\documentclass[twocolumn]{aastex631}

\usepackage{amsmath}
\usepackage{lipsum}
\usepackage{hyperref}
\usepackage{xcolor}
\usepackage{varwidth}

\newcommand{\Rsun}{$\rm R_\odot$}

\newcommand{\Rearth}{$\rm R_\oplus$}

\newcommand{\galex}{{\it GALEX}}
\newcommand{\hipparcos}{{\it HIPPARCOS}}
\newcommand{\gaia}{{\it Gaia}}
\newcommand{\tess}{{\it TESS}}
\newcommand{\wise}{{\it WISE}}

\received{January 5, 2024}
\revised{February 29, 2024}
\accepted{March 1, 2024}

\submitjournal{AAS Journals}

\shorttitle{Setting the stage for the search for life with HWO}
\shortauthors{Harada et al.}
\graphicspath{{./}{figures/}}

\begin{document}

\title{Setting the stage for the search for life with the Habitable Worlds Observatory: \\ Properties of 164 promising planet survey targets}

\author[0000-0001-5737-1687]{Caleb K.~Harada}
\altaffiliation{NSF Graduate Research Fellow}
\affiliation{Department of Astronomy, 501 Campbell Hall \#3411, University of California, Berkeley, CA 94720, USA}

\author[0000-0001-8189-0233]{Courtney D.~Dressing}
\affiliation{Department of Astronomy, 501 Campbell Hall \#3411, University of California, Berkeley, CA 94720, USA}

\author[0000-0002-7084-0529]{Stephen R.~Kane}
\affiliation{Department of Earth and Planetary Sciences, University of California, Riverside, CA 92521, USA}

\author{Bahareh Adami Ardestani}
\affiliation{Sonoma State University, Rohnert Park, CA 94928, USA}
\affiliation{Department of Astronomy, 501 Campbell Hall \#3411, University of California, Berkeley, CA 94720, USA}

\correspondingauthor{Caleb K. Harada}
\email{charada@berkeley.edu}

\begin{abstract}
The Decadal Survey on Astronomy and Astrophysics 2020 (Astro2020) has recommended that NASA realize a large IR/O/UV space telescope optimized for high-contrast imaging and spectroscopy of $\sim$25 exo-Earths and transformative general astrophysics. The NASA Exoplanet Exploration Program (ExEP) has subsequently released a list of 164 nearby ($d<25\,\rm pc$) targets deemed the most accessible to survey for potentially habitable exoplanets with the Habitable Worlds Observatory (HWO). We present a catalog of system properties for the 164 ExEP targets, including 1744 abundance measurements for 14 elements from the Hypatia Catalog and 924 photometry measurements spanning from 151.6~nm to 22~$\mu$m in the \galex, Str\"{o}mgren, Tycho, \gaia, 2MASS, and \wise~bandpasses. We independently derive stellar properties for these systems by modeling their spectral energy distributions with Bayesian model averaging. Additionally, by consulting the literature, we identify \tess~flare rates for 46 stars, optical variability for 78 stars, and X-ray emission for 46 stars in our sample. We discuss our catalog in the context of planet habitability and draw attention to key gaps in our knowledge where precursor science can help to inform HWO mission design trade studies in the near future. Notably, only 33 of the 164 stars in our sample have reliable space-based UV measurements, and only 40 have a mid-IR measurement. We also find that phosphorus, a bio-essential element, has only been measured in 11 of these stars, motivating future abundance surveys. Our catalog is publicly available and we advocate for its use in future studies of promising HWO targets.
\end{abstract}


\section{Introduction} \label{sec:intro}

The Astro2020 Decadal Survey has recommended that NASA devise a large ($\sim$6\,m inscribed diameter), stable, space-based infrared/optical/ultraviolet (IR/O/UV) telescope capable of high-contrast ($10^{-10}$) imaging and spectroscopy \citep{astro2020}. Currently named the Habitable Worlds Observatory (HWO) and targeted to launch in the 2040s, the mission's primary science goals will be to search for biosignatures roughly 25 habitable zone (HZ) exoplanets, while simultaneously enabling transformative general astrophysics \citep{astro2020}.

The Astro2020 report recognized that the unprecedented nature of HWO's science objectives and technology requirements will require sophisticated planning and review leading up to the mission design stage. Astro2020 therefore recommended a Great Observatory Mission and Technology Maturation Program (GOMAP) whose aim is to make early significant investments in the co-maturation of mission concepts and technologies prior to ultimate recommendation and implementation \citep{astro2020}. The outcomes of the GOMAP phase will inform the final design of HWO.

Following the release of Astro2020, the NASA Exoplanet Exploration Program (ExEP) created a Mission Star List\footnote{\url{https://exoplanetarchive.ipac.caltech.edu/docs/2645_NASA_ExEP_Target_List_HWO_Documentation_2023.pdf}} for HWO that included 164 stars ``whose [hypothetical] exo-Earths would be the most accessible for a systematic imaging survey of habitable zones with a 6-m class space telescope in terms of angular separation, planet brightness in reflected light, and planet-star brightness ratio'' \citep{Mamajek_Stapelfeldt_2023}. The ExEP Mission Star List (EMSL) was compiled by analyzing nearby bright stars and adopting inputs from the the mission studies for the Large UV/Optical/IR Surveyor \citep[LUVOIR;][]{LUVOIR_2019} and the Habitable Exoplanet Observatory \citep[HabEx;][]{HabEX_2020}. The broad scope of the list was intended to avoid being overly prescriptive in the required starlight suppression technology or mission requirements \citep{Mamajek_Stapelfeldt_2023}. 

However, while the EMSL contained key observability metrics that were calculated to constrain the sample (e.g., angular separation, contrast ratio between the planet and host star, excess levels of complicating exozodiacal dust, etc.), it lacked other information that may be crucial for identifying a robust sample of potential exo-Earths. For example, the high-energy spectrum of the host star can significantly impact habitability \citep[e.g.,][]{Segura+2003, Rugheimer+2015, Roettenbacher2017}, but the EMSL does not include UV or X-ray radiation measurements. Host star composition, which reflects the composition of the protoplanetary disk, is linked to the outcomes of planet formation and hence may give us hints as to which targets are more or less likely to host habitable rocky planets \citep[e.g.,][]{Gaspar+2016, Santos+2017, Adibekyan+2021, Cabral+2023}. While the EMSL includes the host star metallicity [Fe/H], it is known that the iron abundance alone does not provide a robust proxy for scaling the abundances of other planet-building elements \citep[e.g.,][]{Bitsch+2020, Jorge+2022}.

Starting with the EMSL targets\footnote{We will consider a broader set of stars in future work.}, in this work we construct a catalog of properties of promising nearby targets for the HWO exo-Earth survey. As a step toward building a more complete understanding of these systems and their likelihood of hosting exo-Earths, our catalog contains key properties of these stars that are theorized to be linked to planet habitability. This work, and subsequent precursor science studies, will reduce HWO mission design risk by enabling science and engineering trades that will weigh how key design choices and astrophysical realities may impact the yield of potentially habitable planets characterized by HWO. For example, identifying and characterizing the most promising HWO targets early in mission development will inform critical mission requirements such as inner working angle (IWA), outer working angle (OWA), field of regard, sensitivity, settling time, slew speed, wavelength coverage, and number of filters used for detection and characterization of potentially habitable planets. This work is complementary to the Habitable Worlds Observatory Preliminary Input Catalog \citep[HPIC;][]{Tuchow+2024}, which became public during the revision process of this paper.

The rest of this paper is organized as follows. In Section \ref{sec:data}, we describe the information we have compiled in this catalog, which includes the original EMSL data, photometry from 6 different surveys, 14 elemental abundances, flare rates, stellar variability, and X-ray detections. In Section \ref{sec:analysis}, we describe our analysis of stellar properties derived from SED fitting and the verification of stellar abundances from the literature. We discuss our results and implications of our catalog in Section \ref{sec:discussion}, and conclude in Section \ref{sec:conclusion}.

\section{Data} \label{sec:data}

\subsection{The ExEP Mission Star List}

\begin{figure*}[t!]
    \centering
    \includegraphics[width=\textwidth]{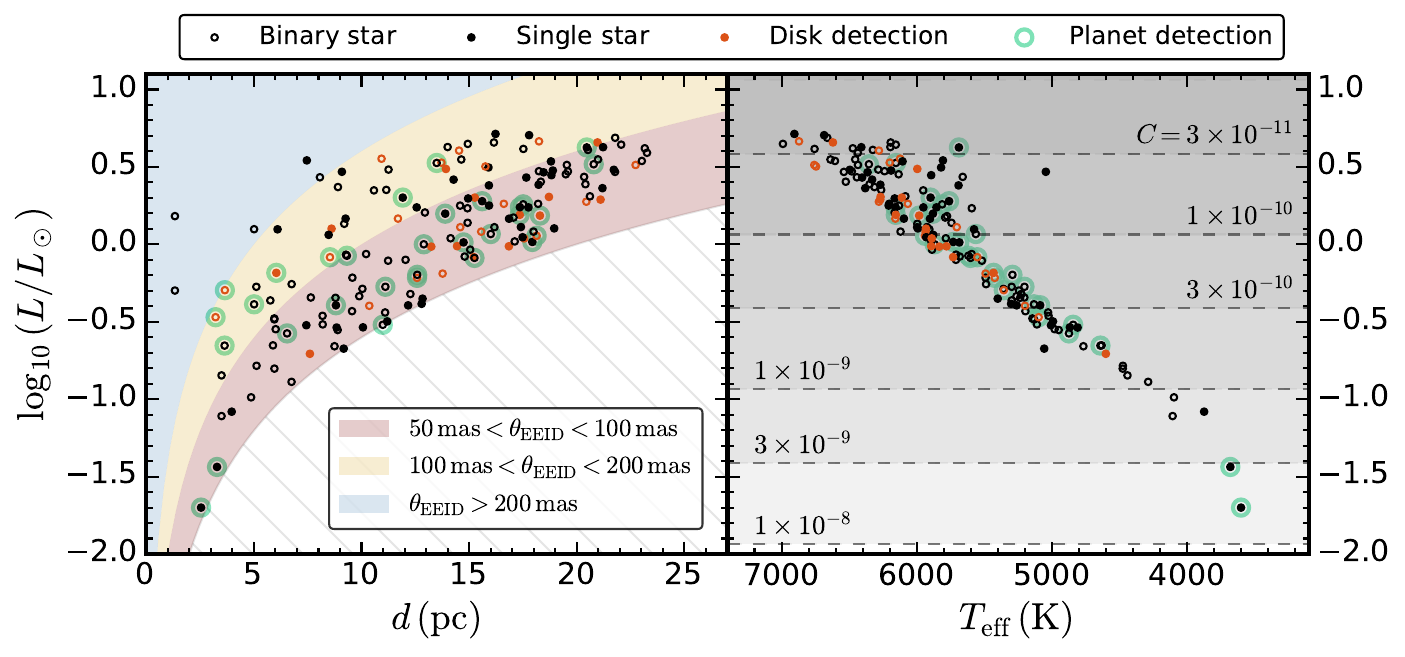}
    \caption{\emph{Left:} EMSL stellar luminosity vs.~distance \citep[data from][]{Mamajek_Stapelfeldt_2023}. The shaded areas show different ranges of the Earth-Equivalent Insolation Distance (EEID) in angular units. \emph{Right:} H-R Diagram for EMSL stars. The shaded intervals show the expected planet-to-star contrast ratio for a {1\,\Rearth} planet at the EEID, assuming a geometric albedo of $p=0.2$ at a phase angle of $\alpha=90^\circ$. In both panels, the closed circles show single stars, while the open circles show stars in binary systems with angular separations $>3\arcsec$. The red circles indicate the presence of a circumstellar debris disk, and the green rings highlight systems known to host at least one planet.}
    \label{fig:emsl_lum_vs_dist_vs_teff}
\end{figure*}

Motivated by the Astro2020 Decadal Survey, \citet{Mamajek_Stapelfeldt_2023} created the ExEP Mission Star List (EMSL) with 164 bright stars whose exo-Earths would be the most accessible for a systematic imaging survey of nearby HZs. To allow flexibility in HWO mission design trade studies and the mission target list, they considered different constraints on the observatory's IWA, exoplanet brightness limits, planet-to-star brightness ratio, the presence of circumstellar debris disks, and stellar multiplicity. Their sample was limited to nearby stars whose HZs would be accessible to a 6-meter-class telescope with starlight suppression technology (capable of contrast down to $2.5 \times 10^{-11}$ in $R_c$ band), and whose HZ terrestrial planets would be bright enough to image and spectroscopically characterize on relatively short timescales \citep[planet magnitudes of $R_c \leq 31$ for integration times up to 60 days;][]{Mamajek_Stapelfeldt_2023}. We briefly summarize their results in this section.

Following the LUVOIR and HabEx mission studies \citep{LUVOIR_2019, HabEX_2020}, \citet{Mamajek_Stapelfeldt_2023} assumed an occurrence rate of terrestrial HZ planets of $\eta_\oplus=0.24$, thus requiring a sample of at least $\sim$100 targets to achieve the Astro2020 goal of imaging 25 exo-Earths. Their calculations adopted the conservative habitable zone limits, defined as the range of semi-major axes between 0.95-1.67\,au for a solar twin for planets between 0.8-1.4\,\Rearth~\citep[e.g.,][]{Kasting+1993, Kopparapu+2013}. For other stars, they scaled the semi-major axis limits by the square root of the solar-normalized stellar luminosity.

Starting from a large initial sample of nearby AFGKM stars, which was descended from the Extreme Precision Radial Velocity (EPRV) Working Group Final Report\footnote{The EPRV Working Group target list was itself drawn from the LUVOIR and HabEx study reports \citep{LUVOIR_2019, HabEX_2020}.} \citep{Crass+2021}, the EMSL was constructed by filling in additional targets from various sources with overlapping spectral types, magnitudes, and distances. For this volume-limited sample of approximately 800 stars, \citet{Mamajek_Stapelfeldt_2023} calculated the angular separation and expected planet brightness assuming 12 different scenarios with various combinations of orbital distance, phase angle, and planet radius. 

In their calculations, \citet{Mamajek_Stapelfeldt_2023} adopted Cousins $R_c$ band photometry because of the $R_c$ filter's overlapping wavelength coverage with the LUVIOR and HabEx bandpasses. Following the LUVOIR and HabEx mission studies, they assumed a planetary geometric albedo of $p=0.2$ for all systems. Then, for each star, they calculated the Earth-equivalent instellation distance (EEID), defined as the orbital separation at which a planet receives the equivalent flux as Earth at 1\,au:
\begin{equation}
    r_{\rm EEID} = 1\,{\rm au}\,\sqrt{L_\star / L_\odot}
\end{equation}
where $L_\star$ is the stellar luminosity\footnote{Note that this basic scaling relation does not account for the detailed spectra of non-solar type stars, which can influence the actual amount of flux received by a planet and hence its temperature.}. Assuming a circular orbit and maximum angular separation, this can also be written in terms of the projected angular separation (in milliarcseconds) as $\theta_{\rm EEID} = r_{\rm EEID} / d_{\rm pc} = r_{\rm EEID} \varpi_{\rm mas}$, where $d_{\rm pc}$ is the distance to the star in parsecs and $\varpi_{\rm mas}$ is the parallax angle in milliarcseconds. 

\citet{Mamajek_Stapelfeldt_2023} then calculated the planet-to-star brightness ratio (i.e., contrast) as
\begin{equation}
    C = F_{\rm p} / F_\star = p \phi(\alpha) (R_{\rm p} / r)^2
\end{equation}
where $p$ is the geometric albedo, $\phi(\alpha)$ is the integral phase function at phase angle $\alpha$ (i.e., the observer-planet-star angle), $R_{\rm p}$ is the planet's radius, and $r$ is the separation between the planet and its host star. They assumed that all planets had circular orbits (such that $r$ is equivalent to the semi-major axis $a$) and adopted a simple Lambertian phase function for isotropic scattering:
\begin{equation}
    \phi(\alpha) = \frac{1}{\pi} \big[ \sin\alpha + (\pi - \alpha)\cos\alpha \big].
\end{equation}

The hypothetical planet's $R_c$ magnitude was then calculated from the contrast ratio as
\begin{equation}
    R_{c,\rm p} = R_{c,\star} - 2.5\log_{10} C
\end{equation}
where $R_{c,\star}$ is the magnitude of the host star.
\begin{deluxetable}{V{1.75cm}V{1.75cm}V{1.75cm}V{1.75cm}}[bt!]
\caption{Summary of EMSL tier criteria \citep[reproduced from][]{Mamajek_Stapelfeldt_2023}. \label{tab:tier_summary}}
\tablehead{
    \colhead{Parameter} &
    \colhead{Tier A} &
    \colhead{Tier B} &
    \colhead{Tier C} 
}
\startdata
Number of stars & 47 & 51 & 66 \\
IWA constraint & 83 mas & 72 mas & 65 mas \\
Planet brightness limit ($R_c$) & 30.5 & 31.0 & 31.0 \\
Planet-to-star contrast limit & $4 \times 10^{-11}$ & $4 \times 10^{-11}$ & $2.5 \times 10^{-11}$ \\
Disk criterion & No known dust disks of any kind. & No disk; or Kuiper Belt disk permitted if $L_{\rm disk}/L_\star \leq 10^{-4}$. & All disks permitted. \\
Treatment of binaries & Single or binary companion at $>10\arcsec$ sep. & Single or binary companion at $5-10\arcsec$ sep. & Single or binary companion at $3-5\arcsec$ sep. \\
\enddata
\end{deluxetable}
 
The initial $\sim$800 systems were ranked by the number of test scenarios (out of 12) where the hypothetical planet satisfied pre-defined limits on angular separation, contrast ratio, and apparent planet $R_c$ magnitude. These limits were informed by the predicted performance of future direct-imaging technology currently under development for HWO \citep{Mamajek_Stapelfeldt_2023}. Further constraints on the presence of debris disks and stellar multiplicity narrowed down the final EMSL list to 164 targets, which were then assigned to a tier (A, B, or C). The authors considered tier A to be the ``best'' targets for an HWO exo-Earth survey, while tiers B and C contained targets with one or more issues which made them less optimal than tier A. The final EMSL contained 47 stars in tier A, 61 in tier B, and 66 in tier C. The criteria defined by \citet{Mamajek_Stapelfeldt_2023} for each tier are summarized in Table \ref{tab:tier_summary}, and we refer the reader to the original ExEP report for more detailed discussion of the selection criteria.

\begin{figure}[t!]
    \centering
    \includegraphics[width=0.45\textwidth]{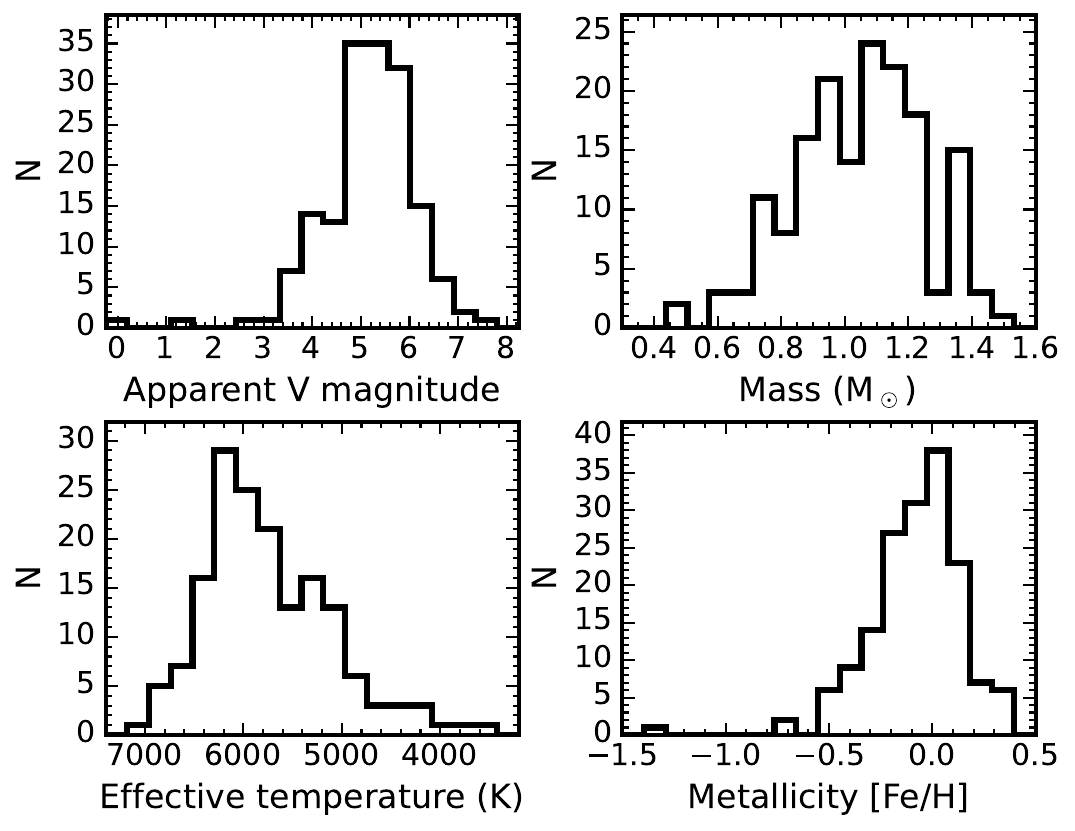}
    \caption{Histograms of the apparent $V$ magnitudes, stellar masses, effective temperatures, and metallicities of EMSL stars \citep[data from][]{Mamajek_Stapelfeldt_2023}.}
    \label{fig:emsl_histograms}
\end{figure}

We accessed the most up-to-date version of the EMSL\footnote{Last accessed: 2024-02-07; \url{https://exoplanetarchive.ipac.caltech.edu/cgi-bin/TblView/nph-tblView?app=ExoTbls&config=DI_STARS_EXEP}} via the NASA Exoplanet Archive Table Access Protocol (TAP). In this work, we include each of the EMSL table columns as they were originally reported, with some minor corrections. Appendix \ref{app:alterations} lists minor inconsistencies that we identified and subsequently corrected in our catalog (e.g., the EMSL table that we downloaded contained an empty column that we remove in our catalog).

\begin{figure}[t!]
    \centering
    \includegraphics[width=0.45\textwidth]{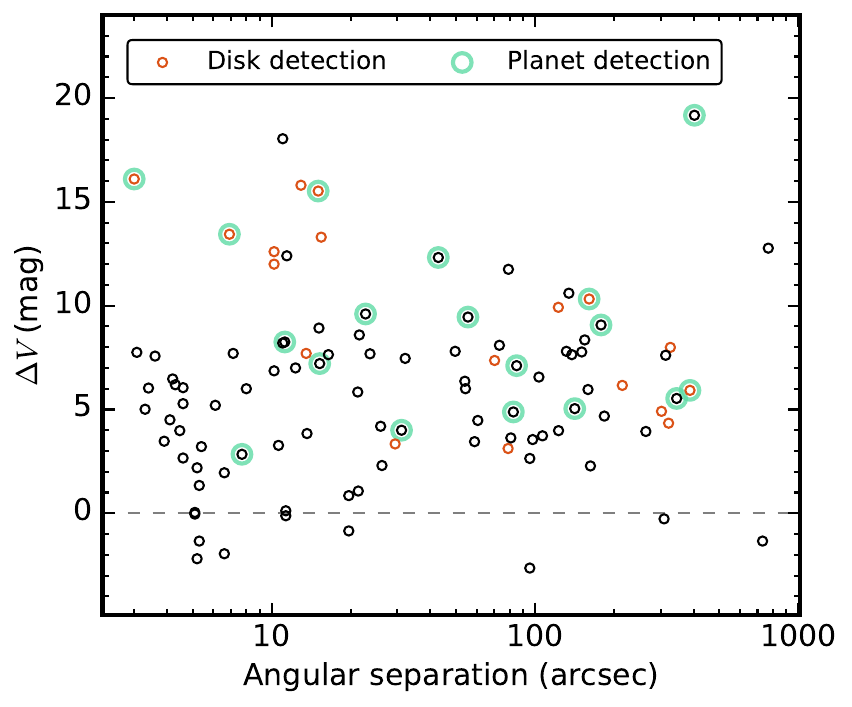}
    \caption{Angular separation vs.~difference in apparent $V$ magnitude for EMSL binary star systems \citep[data from][]{Mamajek_Stapelfeldt_2023}. As in Figure \ref{fig:emsl_lum_vs_dist_vs_teff}, the red circles show systems with a debris disk, and the green rings indicate systems with at least one known planet.}
    \label{fig:emsl_binaries}
\end{figure}

We plot the full sample of EMSL stars in the luminosity-distance plane and in the H-R diagram in Figure \ref{fig:emsl_lum_vs_dist_vs_teff}. Figure \ref{fig:emsl_histograms} shows histograms of the apparent $V$ magnitude, stellar mass, effective temperature, and metallicity of the EMSL sample. Figure \ref{fig:emsl_binaries} demonstrates the range of angular separations and magnitude differences between binary stars across the sample. All values plotted in Figures \ref{fig:emsl_lum_vs_dist_vs_teff}, \ref{fig:emsl_histograms}, and \ref{fig:emsl_binaries} are taken from the original EMSL \citep{Mamajek_Stapelfeldt_2023}.

In the following sections, we expand upon the original EMSL by adding columns with additional system properties. \textit{A summary of all column names, descriptions, units, and references in our catalog (and the EMSL) are provided in Table \ref{tab:columns} in Appendix \ref{app:extras}.}

\subsection{Photometry} \label{subsec:fluxes}

For each of the 164 EMSL stars we searched for archival photometry spanning 151.6\,$\rm nm$ to 22\,$\rm \mu m$. In order to accurately crossmatch the sources in the EMSL table with various photometric catalogs, we first crossmatched the EMSL stars with the \tess~Input Catalog\footnote{Accessed from the Mikulski Archive for Space Telescopes (MAST) via \texttt{astroquery} \citep{astropy_2013_A&A,astropy_2018_AJ}.} \citep[TIC;][]{Stassun+2019} using the TIC ID provided in the EMSL. From the TIC, we obtained \gaia~DR2 IDs, which were then crossmatched to \gaia~DR3 IDs using the \gaia~DR3 \texttt{dr2\_neighbourhood} crossmatch table\footnote{\url{https://gaia.aip.de/metadata/gaiaedr3/dr2_neighbourhood/}} \citep{GaiaCollaboration+2023-AA674A1G,https://doi.org/10.17876/gaia/edr.3/4}. For sources without a \gaia~DR2 ID, we used the \hipparcos~ID provided in the EMSL to crossmatch the sources with \gaia~DR3. We identified \gaia~DR3 IDs for all sources except for $\alpha$~Cen~A, which is the brightest star in our sample ($V\sim0$) and too bright for \gaia.

\subsubsection{\gaia~DR3}

Using the crossmatched \gaia~DR3 IDs, we obtained \gaia~optical photometry and associated uncertainties in the $G$ ($330-1050\,\rm nm$), $G_{BP}$ ($330-680\,\rm nm$), and $G_{RP}$ ($630-1050\,\rm nm$) bandpasses from the \gaia~DR3 \texttt{gaia\_source} table\footnote{\url{https://gaia.aip.de/metadata/gaiaedr3/gaia_source/}} \citep{GaiaCollaboration+2016-AA595A1G, GaiaCollaboration+2023-AA674A1G}, which we accessed via VizieR\footnote{\url{https://vizier.cds.unistra.fr/viz-bin/VizieR}} \citep{gaia_collaboration_catalog_2022}. Due to quality issues that we attribute to detector saturation, we do not include \gaia~photometry for $\alpha$~Cen~A/B.

\subsubsection{AllWISE}

We obtained mid-infrared photometry from the Wide-field Infrared Survey Explorer \citep[\wise;][]{Wright+2010} in four bandpasses: $W_1$ (3.4\,$\mu \rm m$), $W_2$ (4.6\,$\mu \rm m$), $W_3$ (12\,$\mu \rm m$), and $W_4$ (22\,$\mu \rm m$). Using the \gaia~DR3 \texttt{allwise\_best\_neighbour} table\footnote{\url{https://gaia.aip.de/metadata/gaiaedr3/allwise_best_neighbour/}} \citep{GaiaCollaboration+2023-AA674A1G, https://doi.org/10.17876/gaia/edr.3/29}, we crossmatched the \gaia~DR3 IDs of our target stars to their respective identifiers in the AllWISE Catalog, which we accessed using VizieR \citep{Cutri+2014}, and collected the photometry and associated uncertainties for each source where it was available. We do not include any photometry reported to be blended with extended sources, or otherwise flagged as contaminated or poor quality (SNR$<2$). We find measurements satisfying these criteria for 17, 19, 33, and 40 stars in the $W_1$, $W_2$, $W_3$, and $W_4$ bandpasses, respectively. 

\subsubsection{2MASS}

We followed a similar procedure to acquire near-infrared photometry. We used the \gaia~DR3 \texttt{tmass\_psc\_xsc\_best\_neighbour} table\footnote{\url{https://gaia.aip.de/metadata/gaiaedr3/tmass_psc_xsc_best_neighbour/}} \citep{GaiaCollaboration+2023-AA674A1G, https://doi.org/10.17876/gaia/edr.3/40} to crossmatch the \gaia~DR3 IDs of our target stars to their respective The Two Micron All Sky Survey \citep[2MASS;][]{Skrutskie+2006} identifiers. We then used the 2MASS IDs to obtain $J$ (1.24\,$\mu \rm m$), $H$ (1.66\,$\mu \rm m$), and $K_s$ (2.16\,$\mu \rm m$) photometry and associated uncertainties from the 2MASS All-Sky Catalog of Point Sources on VizieR \citep{Cutri+2003}. We excluded any data with an extended source contamination flag or artifact contamination and confusion flag. We also removed any bad photometry indicated by a photometric quality flag of ``X,'' ``U,'' ``F,'' or ``E.'' This resulted in $J$ band photometry for 130 stars, $H$ band photometry for 127 stars, and $K_s$ band photometry for 95 stars.

\begin{figure*}[t!]
    \centering
    \includegraphics[width=\textwidth]{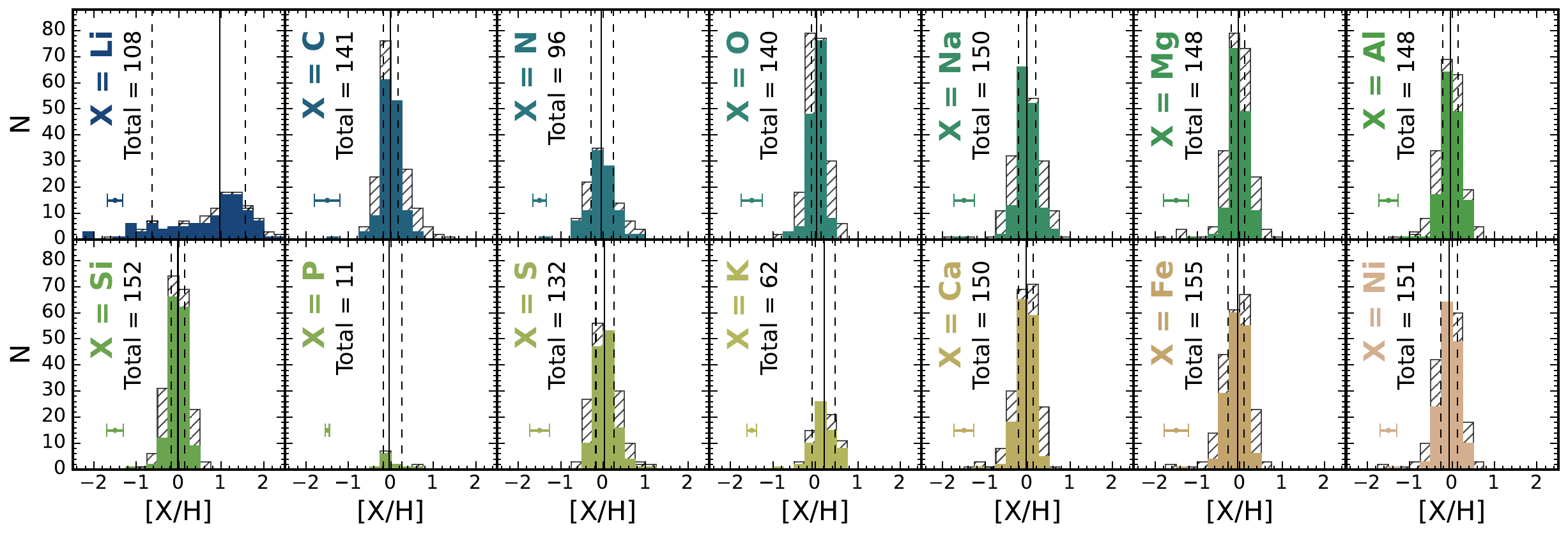}
    \caption{Histograms of the 14 stellar abundances retrieved from the Hypatia Catalog \citep{Hinkel+2014_AJ}. For stars with multiple measurements of the same element, the shaded histograms are computed assuming the mean abundance value and the gray hatched histograms are computed assuming the minimum and maximum abundance values. The element name and corresponding number of stars with at least one abundance measurement for that element are labeled in each panel, and a representative errorbar showing the average difference between the minimum and maximum abundance values is shown in the lower left-hand corner. The vertical lines show the sample median values, and the dashed lines show the 16th and 84th percentiles. Note that the lithium abundances show the greatest amount of spread across the sample due to the strong dependence of [Li/H] on stellar age.}
    \label{fig:hypatia_abundance_histograms}
\end{figure*}

\subsubsection{Tycho-2}

In the optical, we added space-based photometry from the Tycho-2 Catalogue of the 2.5 Million Brightest Stars \citep{Hog+2000}, with measurements from the ESA \hipparcos~satellite's Star Mapper. We crossmatched the \gaia~DR3 IDs of our stars to their Tycho-2 identifiers using the \gaia~DR3 \texttt{tycho2tdsc\_merge\_best\_neighbour} table\footnote{\url{https://gaia.aip.de/metadata/gaiaedr3/tycho2tdsc_merge_best_neighbour/}} \citep{GaiaCollaboration+2023-AA674A1G, https://doi.org/10.17876/gaia/edr.3/16}, then obtained $B_T$ (430\,nm) and $V_T$ (530\,nm) photometry and associated uncertainties from the Tycho-2 Catalog on VizieR using the crossmatched IDs \citep{2000yCat.1259....0H}. We found both $B_T$ and $V_T$ photometry for a total of 132 stars in our sample.

\subsubsection{Str{\"o}mgren}

We obtained additional ground-based UV/optical photometry from the Str\"{o}mgren-Crawford $uvby\beta$ Photometry Catalog \citep{Paunzen+2015} on VizieR \citep{https://doi.org/10.26093/cds/vizier.35800023}. We crossmatched our sources with the \citet{Paunzen+2015} Catalog using their Tycho-2 IDs identified in the previous section. We collected photometry and associated uncertainties in the Str\"{o}mgren $u$ (350\,nm), $v$ (411\,nm), $b$ (467\,nm), and $y$ (547\,nm) bandpasses for 130 of our catalog stars.

\subsubsection{\galex}

Finally, we obtained UV photometry from the Galaxy Evolution Explorer \citep[\galex;][]{Martin+2005}, which includes two bands at 135-175\,nm (FUV) and 175-275\,nm (NUV). We used the \texttt{astroquery} Xmatch service \citep{astropy_2013_A&A, astropy_2018_AJ} to crossmatch the \gaia~DR3 coordinates of our target stars with any \galex~sources within a 30\arcsec~radius in the Revised Catalog of \galex~Ultraviolet Sources \citep{Bianchi+2017}. We collected the FUV and NUV magnitudes with their respective uncertainties of the nearest neighbor \galex~source, and excluded any measurements whose associated artifact or extraction flags were $>0$. We report reliable \galex~measurements for 27 stars in the FUV and 12 stars in the NUV.

\subsection{Stellar abundances} \label{subsec:stellar_abundances}

The Hypatia Catalog is a publicly available spectroscopic abundance catalog consisting of data from 84 literature sources for 50 elements across 3058 stars within 150\,pc of the Sun \citep{Hinkel+2014_AJ}. We obtained stellar abundances for 14 elements from the Hypatia Catalog, including key terrestrial planet building blocks (e.g., Si, Mg, Al, Fe, and Ni) as well as elements that are essential for life on Earth (e.g., C, N, O, P, and S). All of the retrieved abundances were normalized assuming solar values from \citet{Asplund+2009}.

We queried the catalog using the Hypatia API\footnote{Last Accessed: 2024-02-07, \url{https://www.hypatiacatalog.com/api}}, first attempting to crossmatch sources with the Tycho-2 IDs. If the Tycho-2 ID initially failed to return a match, we iteratively reattempted the query using the 2MASS ID, HD name, common name, and finally the Gaia DR2/DR3 ID. For each successful match, we collected all available abundance measurements for the elements listed in Table \ref{tab:abundance_summary}, and recorded the mean, minimum, maximum, and standard deviation of the reported values. In cases where only a single abundance measurement was reported for a given element, we assigned an uncertainty equivalent to the median abundance uncertainty of all the other stars in the sample. For completeness, we also include the total number of abundance values per element for each star in our catalog, in addition to the original literature references for the minimum and maximum abundance values.

A summary of the retrieved abundances is provided in Table \ref{tab:abundance_summary}. Histograms of the mean, minimum, and maximum abundance values are shown in Figure \ref{fig:hypatia_abundance_histograms}.

\subsection{Stellar flare rates} \label{subsec:activity}

We include estimated flare rates for our sample based on light curve observations from the Transiting Exoplanet Survey Satellite (\tess) Mission \citep{Ricker+2014}. Using data from the first 39 sectors of \tess~observations, \citet{Pietras+2022} presented a statistical analysis and extensive catalog of stellar flare events identified in the 2-minute cadence light curves of 330,000 stars using an automated detection pipeline called \texttt{WARPFINDER}. We used the TIC identifiers of our targets to crossmatch our catalog with the flare events reported in the \citet{Pietras+2022} flare catalog. 

For each crossmatched source, we recorded the number of flares detected (in the first 39 sectors of \tess~observations), in addition to the relative amplitude and estimated total energy released by the most energetic flare. Then, using the \texttt{lightkurve} package \citep{lightkurve} to access all two-minute cadence \tess~observations of each target on MAST, we determined the number of \tess~sectors with observations and how many of these sectors were observed up to sector 39. We note that 148 of our stars have at least one sector of \tess~observations, and the median number of \tess~sectors observed per star is 3. Using this information, combined with the known duration of each sector \citep[27.4 days;][]{Ricker+2014}, we calculated the stellar flare rate (in $\rm s^{-1}$) for each star with at least one flare detected by \citet{Pietras+2022}. We identified 44 such stars in our sample.

\begin{figure}[t!]
    \centering
    \includegraphics[width=0.45\textwidth]{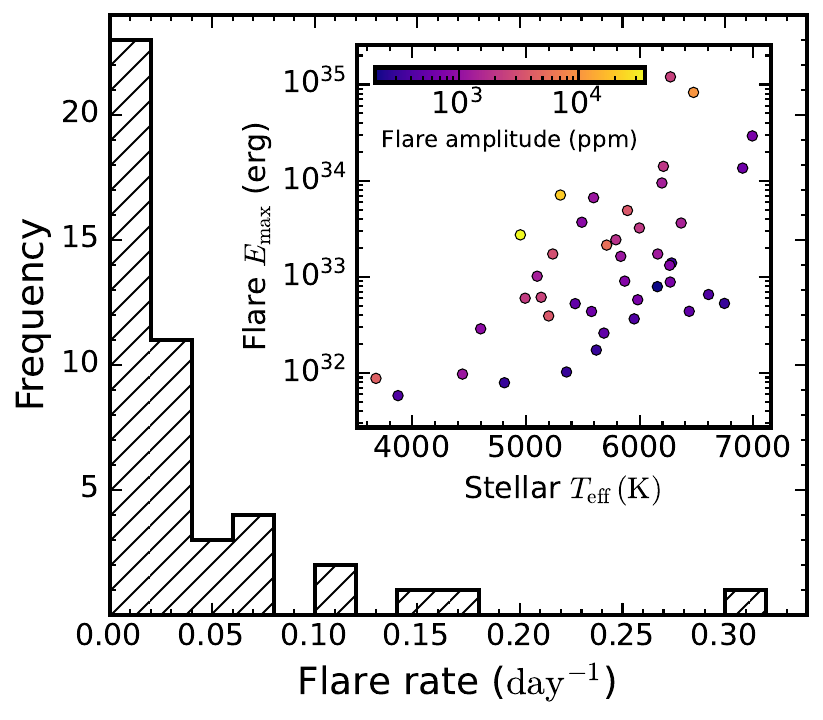}
    \caption{Histogram of stellar flare rates from \tess, calculated using the results of \citet{Pietras+2022} and \citet{Yang+2023}. The inset plot shows the estimated energy released by the most energetic flare as a function of stellar effective temperature. The colorbar indicates the measured relative flare amplitude.}
    \label{fig:flare_rate}
\end{figure}

For completeness, we also checked for flares using the \citet{Yang+2023} catalog of stellar flares from \tess. Using data from cycles 1-30 of \tess, \citet{Yang+2023} independently measured 60,810 flare events from 13,478 stars. We identified two additional flare sources in the \citet{Yang+2023} catalog that were not found in the \citet{Pietras+2022} sample of 44 stars: HD~166 and GJ~811. For these two targets, we recorded the number of flares detected in the first 30 sectors of \tess~observations and the relative amplitude and energy released by the most energetic flare \citep{Yang+2023}. We also estimated the flare rate following the procedure described above. A histogram of all 46 \tess~flare rates and a plot of the flare energy as a function of stellar effective temperature are shown in Figure \ref{fig:flare_rate}.

Lastly, we also searched the literature for stellar flare information at higher-energy wavelengths. We identified 7 systems in our catalog with X-ray flare detections. These X-ray flares were identified in the XMM-Newton serendipitous source catalogue \citep{Pye+2015} for 61~Cyg~A/B, 70~Oph~A/B, $\xi$~Boo~A, HD~95735, 3~UMa, $\epsilon$~Eri, and $\kappa$1~Cet. In our catalog, we include an X-ray flare detection flag for these sources. 
\begin{deluxetable}{cCCCCC}[bt!]
\caption{Summary of stellar abundances retrieved from the Hypatia catalog \citep{Hinkel+2014_AJ}. \label{tab:abundance_summary}}
\tablehead{
    \colhead{Element} &
    \colhead{\# Stars} &
    \colhead{\# Meas.} &
    \colhead{Min.} &
    \colhead{Median} &
    \colhead{Max.} \\
    \colhead{} &
    \colhead{} &
    \colhead{} &
    \colhead{{(dex)}} &
    \colhead{{(dex)}} &
    \colhead{{(dex)}}
}
\startdata
{[Li/H]} & 108 & 361 & -2.11 & 0.97 & 2.37 \\
{[C/H]} & 141 & 803 & -1.27 & -0.02 & 0.66 \\
{[N/H]} & 96 & 165 & -1.41 & -0.04 & 0.76 \\
{[O/H]} & 140 & 928 & -0.71 & 0.02 & 0.38 \\
{[Na/H]} & 150 & 1200 & -1.58 & -0.02 & 0.56 \\
{[Mg/H]} & 148 & 1143 & -1.13 & -0.04 & 0.42 \\
{[Al/H]} & 148 & 1030 & -1.22 & -0.04 & 0.49 \\
{[Si/H]} & 152 & 1438 & -1.1 & -0.02 & 0.43 \\
{[P/H]} & 11 & 17 & -0.28 & -0.04 & 0.54 \\
{[S/H]} & 132 & 504 & -0.46 & 0.03 & 1.02 \\
{[K/H]} & 62 & 88 & -0.81 & 0.2 & 0.62 \\
{[Ca/H]} & 150 & 1192 & -1.09 & -0.03 & 0.36 \\
{[Fe/H]} & 155 & 2811 & -1.3 & -0.05 & 0.39 \\
{[Ni/H]} & 151 & 1381 & -1.38 & -0.06 & 0.46 \\
\enddata
\end{deluxetable} 

\subsection{Stellar variability} \label{subsec:variability}

We obtained stellar variability information for our stars using the Variability Catalog of Stars Observed during the \tess~Prime Mission \citep{Fetherolf+2023}. This catalog contains over 84,000 periodic variables (with periods ranging from 0.01-13 days) identified with high or moderate confidence, including $\sim$65,000 variable stars that were not previously identified in the literature \citep{Fetherolf+2023}.

We used the TIC identifiers to crossmatch our targets with the \citet{Fetherolf+2023} variability catalog, which we downloaded from the MAST \citep{https://doi.org/10.17909/f8pz-vj63}. For each crossmatched source, we collected the type of variability solution \citep[i.e., single or double sinusoid or autocorrelation function;][]{Fetherolf+2023} in addition to the variability period(s) and amplitude(s). The single sinusoid periodic signals can indicate a variety of stellar activity, including rotational variations caused by starspot activity, a dominant pulsation mode, or ellipsoidal variations in stellar binaries \citep{Fetherolf+2023}. On the other hand, double sinusoid solutions are preferred for stars with multiple dominant periodicities, such as multimodal pulsations or differential stellar rotation, while autocorrelation function (ACF) solutions typically represent short-period eclipsing binary systems or periodic non-sinusoidal rotational variability \citep{Fetherolf+2023}. 

\begin{figure}[t!]
    \centering
    \includegraphics[width=0.45\textwidth]{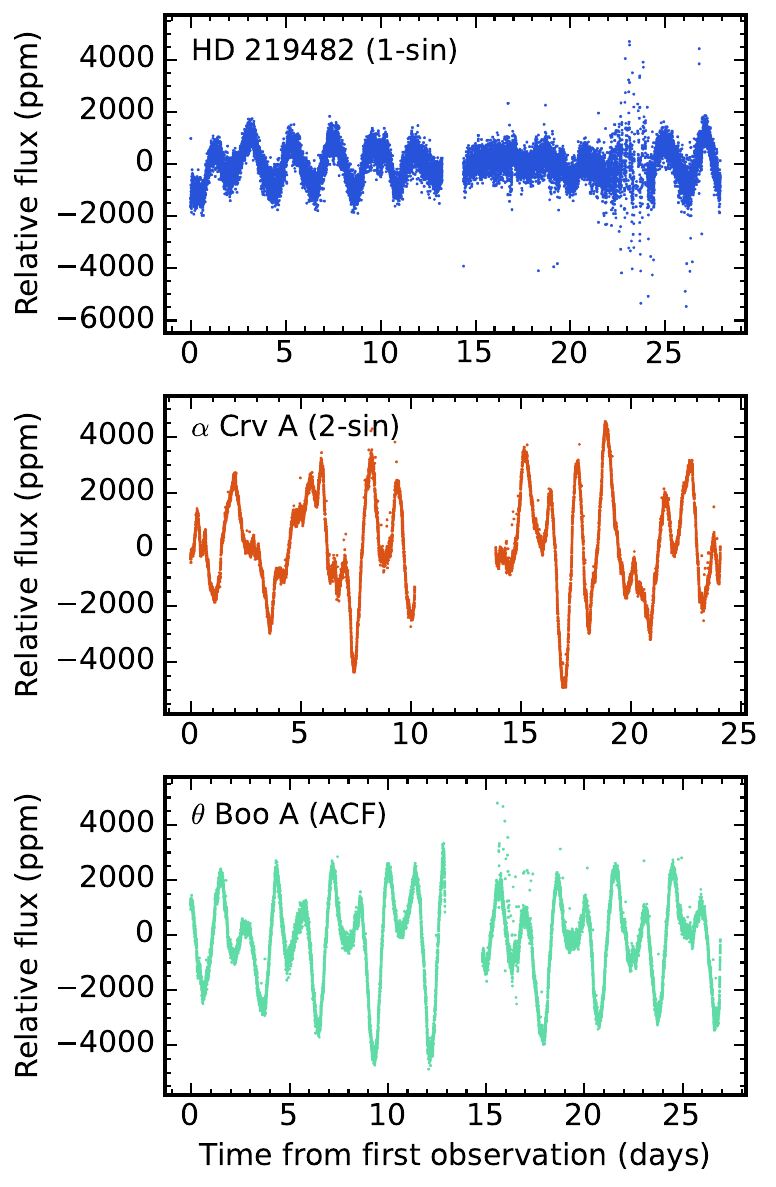}
    \caption{Example \tess~light curves for variable stars in our sample identified by \citet{Fetherolf+2023}. The top panel shows an example of a single sinusoid solution for HD~219482, the middle panel shows a double sinusoid solution for $\alpha$~Crv~A, and the bottom panel shows an ACF solution for $\theta$~Boo~A. The data used to create this figure were downloaded from MAST using the \texttt{Lightkurve} package \citep{lightkurve}.}
    \label{fig:variability_examples}
\end{figure}

\begin{figure}[t!]
    \centering
    \includegraphics[width=0.45\textwidth]{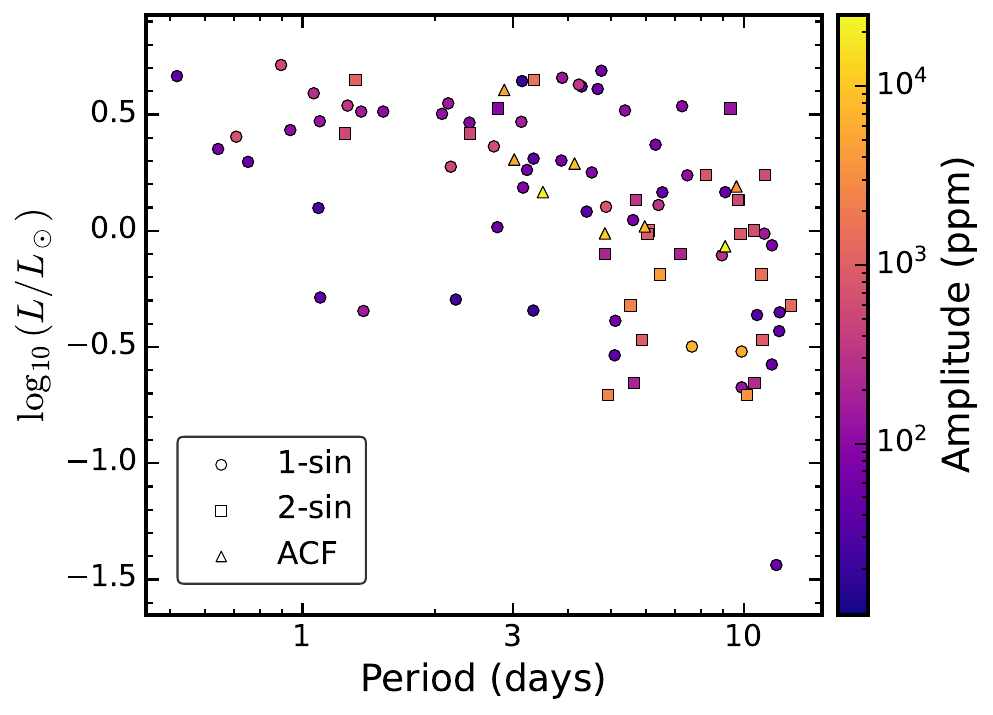}
    \caption{Stellar luminosity vs.~variability period for variable stars identified in the \citet{Fetherolf+2023} catalog. The colorbar indicates the relative variability amplitude, and the variability solution type is indicated by the marker shape.}
    \label{fig:variability_demographics}
\end{figure}

We found 78 stars in our sample with an indication of variability in the \tess~photometry. Figure \ref{fig:variability_examples} shows examples of \tess~light curves, which we downloaded from the MAST using the \texttt{Lightkurve} package \citep{lightkurve}, for stars with different variability solutions identified by \citet{Fetherolf+2023}. We plot the luminosity of these stars as a function of their variability period in Figure \ref{fig:variability_demographics}.

For completeness, and to search for variable sources with periods longer than 13 days, we also queried the American Association of Variable Star Observers (AAVSO) International Variable Star Index (VSX). This comprehensive variability catalog combines variable star observations from multiple published sources utilizing both ground- and space-based photometry\footnote{\url{https://www.aavso.org/vsx/index.php?view=about.top}}. We successfully crossmatched 65 of our stars with the VSX catalog classified as either ``variable'' or ``suspected variable,'' of which 30 were not included in the \citet{Fetherolf+2023} \tess~catalog. For each of these 65 stars, we provide a URL link to the respective VSX entry in our catalog.

\subsection{X-ray emission} \label{subsec:xrays}

Finally, we identified sources in our catalog with previous X-ray detections by crossmatching our sample with the Millions of Optical-Radio/X-ray Associations (MORX) Catalogue \citep{Flesch_2016, Flesch_2023}. MORX combines published optical, radio, and X-ray sky catalogs to match probable radio/X-ray associations with optical counterparts for over 3 million sources. The catalog includes X-ray detections from the \textit{Chandra} \citep{Weisskopf+2000}, \textit{XMM-Newton} \citep{Jansen+2001}, \textit{Swift} \citep{Burrows+2005}, and \textit{ROSAT} \citep{Voges+1999} satellite missions.

We used VizieR \citep{2023yCat.5158....0F} and the \texttt{astroquery} Xmatch service \citep{astropy_2013_A&A, astropy_2018_AJ} to crossmatch the \gaia~DR3 coordinates of our targets with any sources identified in the MORX catalog within a 30\arcsec~radius. We collected the \textit{Chandra}, \textit{XMM-Newton}, \textit{Swift}, and/or other identifiers of the best neighbor X-ray source. We verified that each crossmatched source was classified as a star and an X-ray source by checking the corresponding quality flags and requiring that the MORX algorithm was $>80$\% confident in its classification. We identified 41 of our targets in the MORX catalog, bringing the total number of sources in our catalog with X-ray emission up to 46 (including the X-ray flare detections described in Section \ref{subsec:activity}).

\section{Analysis} \label{sec:analysis}

\subsection{Stellar SED fitting} \label{subsec:sed_fitting}

Using the photometry collected from the literature (described in Section \ref{subsec:fluxes}), we constructed spectral energy distributions (SEDs) for each star in our catalog. We then used the open source \texttt{ARIADNE} package \citep{Vines+2022MNRAS} to fit synthetic SED models to the data in order to independently constrain stellar radii ($R_\star$), masses ($M_\star$), effective temperatures ($T_{\rm eff}$), and luminosities ($L_\star$).

\texttt{ARIADNE} leverages Bayesian Model Averaging \citep[BMA; e.g.,][]{Fragoso+2018} and nested sampling techniques \citep[e.g.,][]{Skilling2004, Skilling2006} to accurately predict stellar properties from SEDs. The advantage of BMA is that it addresses systematic biases between different stellar atmosphere models by incorporating information from each one to arrive at a solution. Specifically, \texttt{ARIADNE} incorporates the PHOENIX \citep{Husser+2013}, BT-Settl and BT-Cond \citep{Allard+2012}, BT-NextGen \citep{Hauschildt+1999, Allard+2012}, Kurucz \citep{Kurucz1993}, and CK04 \citep{Castelli+2004} stellar atmosphere models, which vary in their treatment of opacities and their assumptions regarding stellar abundances, local thermodynamic equilibrium, and convection/overshooting. \texttt{ARIADNE} first fits each stellar model independently, then computes the weighted average of the posterior samples of each model using the Bayesian evidence as the weight. The best-fit parameters and uncertainties are then estimated from this weighted average posterior \citep{Vines+2022MNRAS}.

While the \texttt{ARIADNE} code includes a built-in framework to automatically retrieve archival photometry for any given \gaia~star \citep{Vines+2022MNRAS}, we chose instead to override this step and use the photometry we retrieved for our catalog. This choice was made in order to maintain self-consistency, and to avoid potential issues with the \texttt{ARIADNE} crossmatching routines and quality tolerances. Moreover, we found that photometry in certain bandpasses was less suitable for SED fitting due to systematic discrepancies with all of the models. Specifically, we found that including the \galex~FUV/NUV and Str\"{o}mgren $u$ photometry would significantly degrade the quality our fits, resulting in unreliable posterior constraints and often large offsets between the models and data. Therefore, we excluded any \galex~FUV/NUV and Str\"{o}mgren $u$ photometry from our SED analyses. We also note that the current version of \texttt{ARIADNE} at the time of writing does not support fitting the mid-IR \wise~photometry, so these data were also excluded from the analysis.

We selected priors for each fit based on the stellar parameters included in the EMSL, which are summarized in Table \ref{tab:priors_summary}. Given some of the suspiciously narrow uncertainties associated with many of the parameters in the EMSL, we placed lower limits on the allowed variances of the normal priors used for the effective temperature, surface gravity, and metallicity in order to avoid being overly prescriptive with our prior choices. For the variance of the normal distance prior, we used the parallax angle uncertainty reported in the EMSL to calculate a corresponding distance uncertainty (as there was none reported in the EMSL). Finally, for the stellar radius we assumed a uniform prior ranging between half and twice the value of the EMSL stellar radius.

We initially included an extinction term ($A_V$) in our models, but we found that fitting for $A_V$ could in some cases lead to a systematic over-estimate of effective temperature compared to the EMSL values. We attributed this result to the relatively small distances to the stars in our sample ($d < 25 \,\rm pc$), which could imply negligible interstellar extinction for some targets. In these cases, including the superfluous $A_V$ parameter could lead to over-fitting, resulting in the systematic disagreement in $T_{\rm eff}$. To mitigate this issue, we ran two different version of our SED fits: one with an extinction term and one without.

\begin{figure}[t!]
    \centering
    \includegraphics[width=0.45\textwidth]{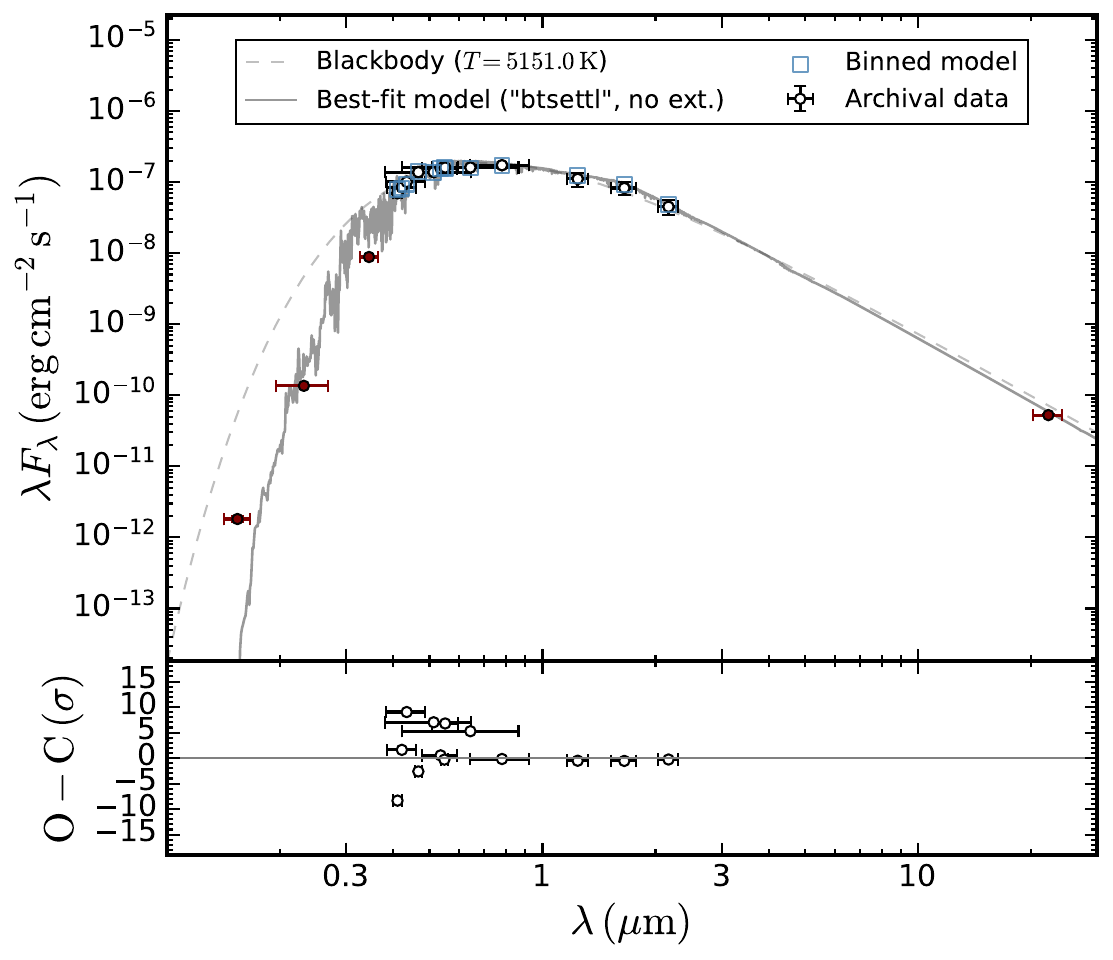}
    \caption{\emph{Top:} Example best-fit SED model for HD~10476, computed with \texttt{ARIADNE}. The points with errorbars show the archival photometry, and the blue squares show the synthetic model photometry. The UV and mid-IR points (shown in maroon) were not used to compute the fit. In this case, the overplotted model with the highest Bayesian evidence was the ``btsettl'' stellar atmosphere with no extinction \citep{Allard+2012}. The parameter posterior distributions derived from this example fit are shown in Appendix \ref{app:extras}. \emph{Bottom:} Observed minus calculated photometry residuals, normalized to the uncertainty the data. The full set of 164 SED plots can be accessed through the online journal.}
    \label{fig:example_sed}
\end{figure}

For both cases, we used the BMA algorithm implemented in \texttt{ARIADNE} \citep{Vines+2022MNRAS} with \texttt{dynesty} nested sampling \citep{Speagle2019} to derive best-fit parameters and posterior distributions for each of our 164 stars. We found that setting the number of live points to 1000 and using a stopping criterion of $d\log z = 0.5$ resulted in robust, converged posteriors. We then compared the Bayesian evidences of the SED models with and without interstellar extinction and selected the best-fit parameters from the model with the higher Bayesian evidence.
\begin{deluxetable}{RL}[bt!]
\caption{Summary of priors used for SED fitting. \label{tab:priors_summary}}
\tablehead{
    \colhead{Parameter} &
    \colhead{Prior}
}
\startdata
T_{\rm eff} \,(\rm K) & \mathcal{N}\big(T_{\rm eff,EMSL},\, \max(\sigma_{T_{\rm eff,EMSL}}, 100\,{\rm K})\big) \\
\log g \,(\rm cgs) & \mathcal{N}\big(\log (g)_{\rm EMSL},\, \max(\sigma_{\log (g)_{\rm EMSL}}, 0.1)\big) \\
{\rm [Fe/H]} & \mathcal{N}\big({\rm [Fe/H]}_{\rm EMSL},\, \max(\sigma_{{\rm [Fe/H]}_{\rm EMSL}}, 0.05)\big) \\
d \,(\rm pc) & \mathcal{N}\big({1\arcsec}/{\varpi_{\rm EMSL}},\, {\sigma_{\varpi_{\rm EMSL}}}/{\varpi_{\rm EMSL}^2} \big) \\
R_\star \,(\rm R_\odot) & \mathcal{U}\big(0.5 R_{\star, \rm EMSL},\, 2 R_{\star, \rm EMSL}\big) \\
A_V \,(\rm mag)\tablenotemark{1} & \mathcal{U}\big(0, A_{V, \rm max}\big) \\
\enddata
\tablecomments{$\mathcal{N}(\mu, \sigma)$ represents a normal distribution with mean $\mu$ and variance $\sigma^2$; $\mathcal{U}(a, b)$ represents a uniform distribution bounded between $a$ and $b$.}
\tablenotetext{1}{Only used for fits where $A_V$ was used as a free parameter. $A_{V, \rm max}$ is the maximum line-of-sight extinction from the updated SFD galactic dust map \citep{Schlegel+1998, Schlafly+2011}.}
\end{deluxetable} 
An example of the best-fit SED for the star HD~10476 is shown in Figure \ref{fig:example_sed}---note that, for plotting the SED, \texttt{ARIADNE} chooses the model grid that yielded the highest Bayesian evidence, but the best-fit parameters for the plotted model are estimated from the weighted average posteriors from all of the fitted models. The corresponding BMA posterior distributions for each of the fit parameters are shown in Figure \ref{fig:sed_posteriors} in Appendix \ref{app:extras}.

Lastly, we note that \texttt{ARIADNE} also estimates stellar age using the SED fit parameters together with the \texttt{isochrones} package \citep{isochrones} and the MESA Isochrones and Stellar Tracks (MIST) library \citep{MIST}. While we caution that single-star isochrone fitting may be insufficient to reliably estimate stellar ages (as is apparent in the wide age uncertainties resulting from many of the \texttt{ARIADNE} fits), we include the estimated ages and uncertainties in our catalog as a baseline for future work investigating stellar ages more closely.

\begin{figure*}[t!]
    \centering
    \begin{minipage}{0.49\textwidth}
        \includegraphics[width=\textwidth]{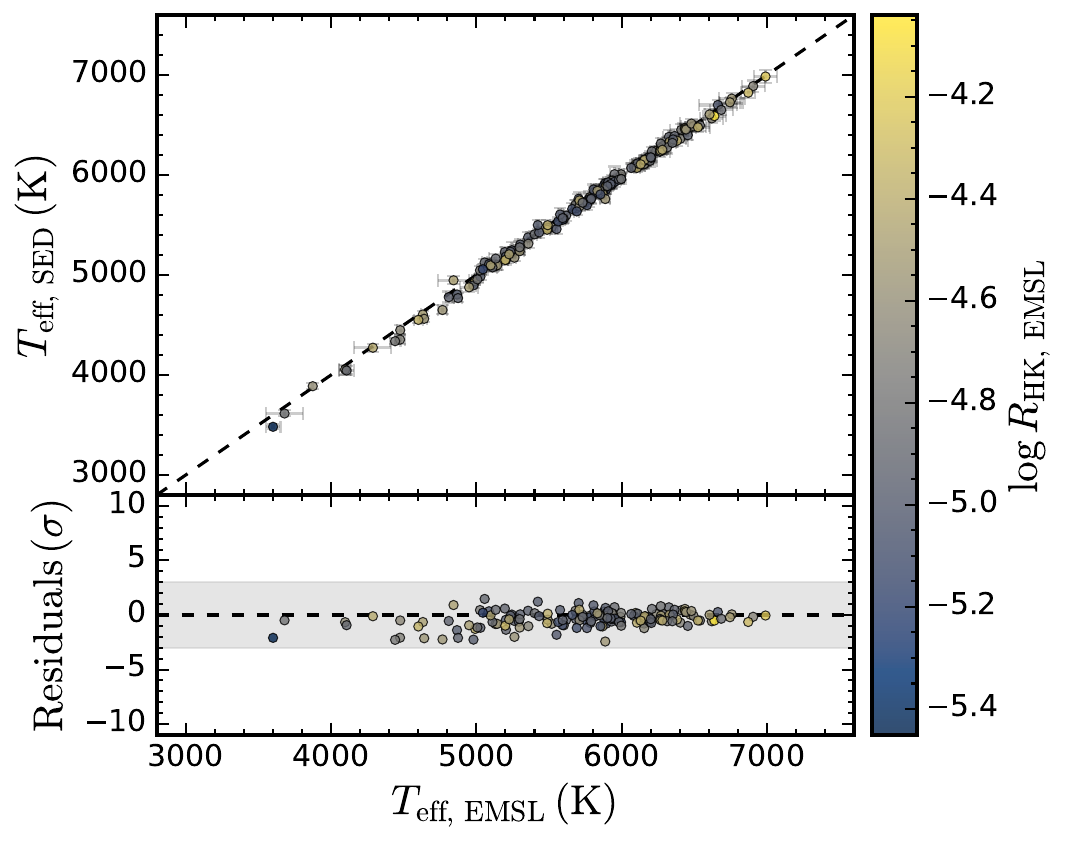}
    \end{minipage}
    \begin{minipage}{0.49\textwidth}
        \includegraphics[width=\textwidth]{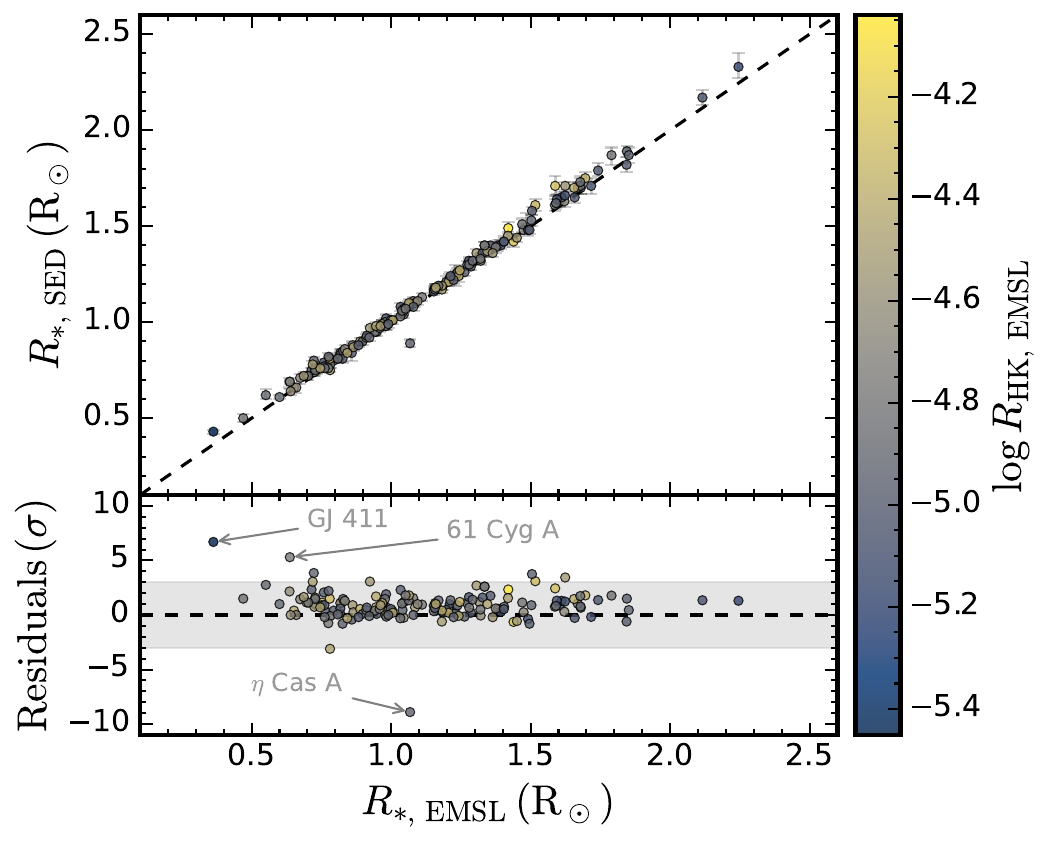}
    \end{minipage}
    \caption{\emph{Left:} Stellar effective temperature ($T_{\rm eff}$) derived from the \texttt{ARIADNE} SED fits vs.~$T_{\rm eff}$ reported in the EMSL table (top), along with the residuals normalized by the quadrature sum of their respective uncertainties (bottom). The black dashed line in the top panel shows a 1-to-1 correspondence, and the gray shaded box in the bottom panel highlights the region of $<3\sigma$ discrepancy. \emph{Right:} Same as the left-hand side, but for stellar radius ($R_*$).}
    \label{fig:ariadne_one_to_one}
\end{figure*}

In general, we found that our SED analyses resulted in stellar parameters that were consistent with the EMSL, which itself adopted values from various literature sources \citep{Mamajek_Stapelfeldt_2023}. Figure \ref{fig:ariadne_one_to_one} shows a comparison between the $T_{\rm eff}$ and $R_\star$ derived from our SED fits and those reported in the EMSL. We found that the SED-derived effective temperatures are all consistent with the EMSL values within $3\sigma$, with a residual RMS of 41.8\,K. For the stellar radii, we found that 94\% of the SED-derived values were consistent with the EMSL within $3\sigma$, achieving a residual RMS of 0.035\,\Rsun. Further discussion of these results is presented in Section \ref{subsec:stellar_properties}

\subsection{Stellar metallicity verification} \label{subsec:met_verification}

While the original EMSL included stellar metallicity measurements for each star, the assumed values were selected from disparate literature sources \citep{Mamajek_Stapelfeldt_2023}. This introduced the possibility of systematic biases that depend on the particular method used to determine the stellar metallicity by each separate source catalog. The Hypatia catalog \citep{Hinkel+2014_AJ} allows for more robust stellar metallicity estimates because it combines multiple values from different source catalogs. These can subsequently be averaged together to yield more statically robust metallicity estimates.

We checked for consistency between the [Fe/H] metallicities reported in the EMSL and the average iron abundances from the Hypatia catalog. The values of [Fe/H] from the EMSL and Hypatia are plotted against one another in Figure \ref{fig:metallicity_one_to_one}. We calculated the discrepancy between the EMSL and Hypatia metallicities (in $\sigma$) by taking the quadrature sum of the EMSL metallicity uncertainty and the standard deviation of all Hypatia metallicities. All but one of the stars (GJ 380; see Section \ref{subsec:stellar_compositions} for discussion) had metallicities from the EMSL and Hypatia that were consistent within $2\sigma$. We calculated the RMS of the residuals to be 0.048~dex. 

We also note that the metallicity uncertainties we derived from Hypatia are generally larger than those reported in the EMSL (0.065~dex median for Hypatia compared to 0.02~dex median for the EMSL) due to the systematic differences between the various Hypatia source catalogs (e.g., in stellar models, line lists, instrument resolution, instrument wavelength coverage, etc.). In principle, our more conservative metallicity uncertainties are more reliable because they take into account multiple independent analyses. However, we caution that this means that Figure \ref{fig:metallicity_one_to_one} is not necessarily an apples-to-apples comparison, and that real trends could be blurred out due to non-astrophysical effects. We discuss these results further in Section \ref{subsec:stellar_compositions}.

\section{Discussion} \label{sec:discussion}

\begin{figure}[t!]
    \centering
    \includegraphics[width=0.45\textwidth]{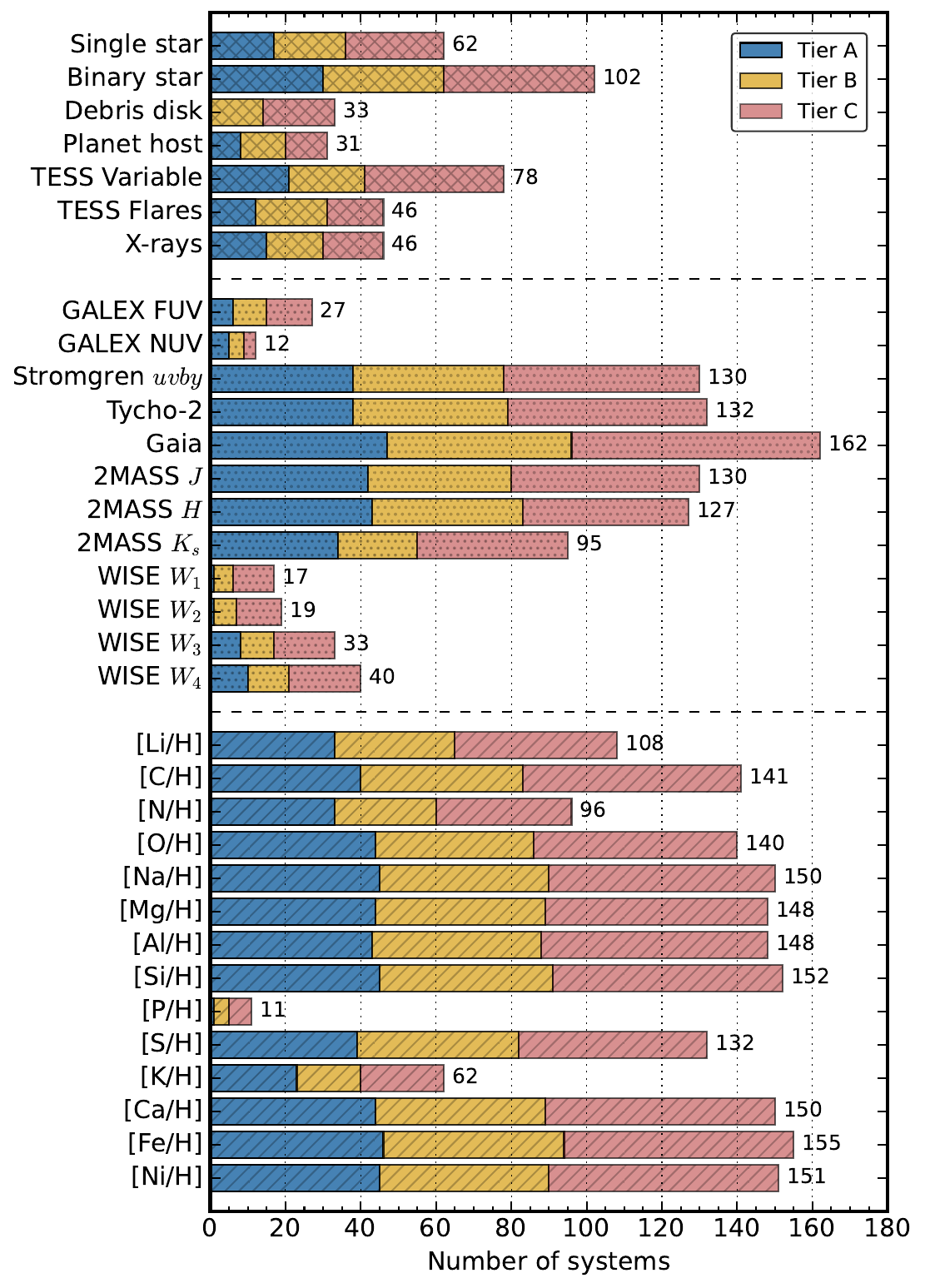}
    \caption{Bar graph showing a summary of the information in our catalog, demonstrating the (in)completeness of our knowledge about these systems. Each attribute is subdivided into the three tier levels defined by \citet{Mamajek_Stapelfeldt_2023}.}
    \label{fig:emsl_barchart}
\end{figure}

\begin{figure*}[t!]
    \begin{interactive}{js}{spores_interactive_demographics.html}
        \centering
        \includegraphics[width=0.85\textwidth]{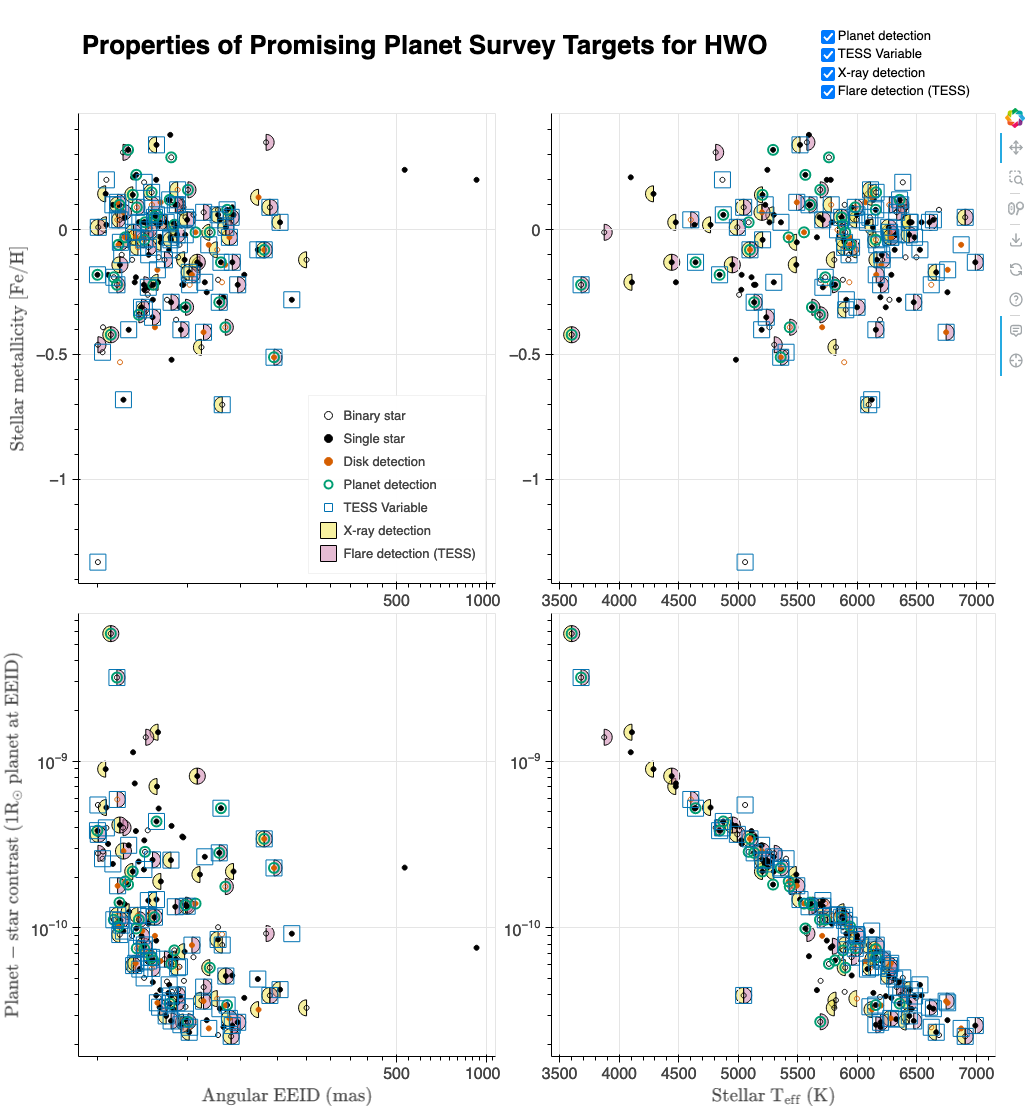}
    \end{interactive}
    \caption{Physical and observable properties of stars in our catalog. Starting from the upper left-hand panel and moving clockwise, the panels show stellar metallicity vs.~angular EEID, stellar metallicity vs.~effective temperature, estimated planet-star contrast ratio (in $R_c$ band) vs.~effective temperature, and estimated planet-star contrast ratio vs.~angular EEID. Different markers indicating stellar multiplicity, the presence of debris disks and planets, stellar variability, and flare and X-ray detections are defined in the legend. An interactive version of this figure can be accessed through the online journal. The Bokeh.org interactive version allows you to highlight different subsets of the catalog, zoom and pan in the different, linked panels, and toggle tooltips and finder crosshairs on or off.}
    \label{fig:emsl_demographics}
\end{figure*}

Our updated catalog of properties of 164 promising targets for HWO includes 924 photometry measurements in 18 different bandpasses spanning from 151.6\,nm to 22\,$\rm \mu m$, 1744 abundance measurements for 14 different elements, variability metrics for 78 stars, X-ray detections for 46 stars, and flare rates for 46 stars. A more granular summary of the catalog properties is presented in Figure \ref{fig:emsl_barchart}, and key physical and observable properties of the stars in our catalog are plotted in Figure \ref{fig:emsl_demographics}. This catalog of stellar properties is intended to provide a starting point for future precursor and preparatory science leading up to HWO's final mission design and eventual launch the 2040s.

\subsection{Photometry and SED fits} \label{subsec:stellar_properties}

We produced SEDs for each of the 164 stars in our sample using photometry drawn from the literature, and fit SED models to independently verify stellar parameters. Using Bayesian model averaging and 6 standard libraries of stellar atmosphere models (see Section \ref{subsec:sed_fitting}), we derived $T_{\rm eff}$, $R_\star$, $M_\star$, and $L_\star$ that were consistent with the literature values reported in the EMSL. A comparison of our SED fit results and literature values for $T_{\rm eff}$ and $R_\star$ are shown in Figure \ref{fig:ariadne_one_to_one}.

We note that the stellar radius we derived for 10 of our stars appear to be discrepant from the values reported in the literature by more than $3\sigma$, with GJ~411, 61~Cyg~A, and $\eta$~Cas~A having the largest disagreement. However, close inspection of the SED fits and posterior distributions did not reveal any fits with unexpectedly poor quality. We therefore attribute these discrepancies to the lack of radius uncertainties reported in the EMSL (possibly because the stellar radius is often a calculated, rather than observed, quantity). We also note that the EMSL also does not report stellar mass uncertainties. Knowing the uncertainties associated with both values would be necessary to confidently determine whether these parameters are in fact significantly discrepant.

Furthermore, we note that the $T_{\rm eff}$ uncertainties from the literature may be underestimated in many cases \citep[see, e.g.,][]{Tayar+2022}. The median uncertainty in $T_{\rm eff}$ reported in the EMSL is 20\,K, which is more than a factor of two smaller than the median uncertainty we derived from SED fitting, about 47\,K. Our stellar luminosity uncertainties are also about a factor of two larger than the literature values. For these reasons, we advocate that future studies of these stars adopt the slightly more conservative stellar parameters from our uniform SED fits.  

\subsection{Stellar SEDs and planetary habitability} \label{subsec:sed_habitability}

Planetary habitability is critically linked to the energy output of the host star \citep{Buccino+2007}. The EMSL \citep{Mamajek_Stapelfeldt_2023} was compiled assuming the classical HZ limits \citep[i.e., water loss and maximum greenhouse;][]{Kasting+1993, Kopparapu+2013, Kane2016, Hill2023}, which were scaled by the square root of the stellar luminosity for each star. While this is a useful first approximation, the sensitivity of exoplanet habitability to the spectral energy distribution of the host star is more complex, especially for non-solar type stars which have fundamentally different SEDs. For example, two stars with the same bolometric flux will yield different effective temperatures if their SEDs are different.

\begin{figure}[t!]
    \centering
    \includegraphics[width=0.49\textwidth]{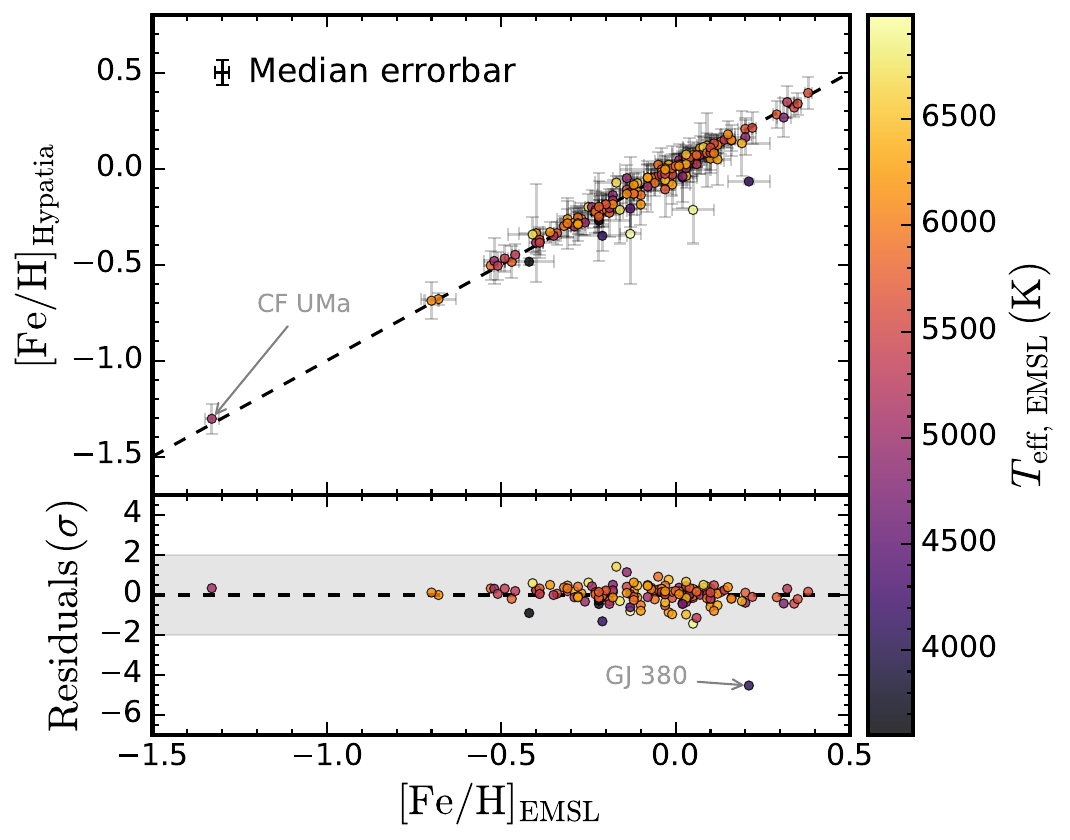}
    \caption{(Top:) Median iron abundance metallicity [Fe/H] from the Hypatia Catalog vs.~the [Fe/H] reported in the original EMSL table, with a 1-to-1 correspondence shown by the black dashed line. The stellar effective temperature from original EMSL table is indicated by color. A representative $1\sigma$ median errorbar is shown in the upper left-hand corner. (Bottom:) The residuals between the median Hypatia and EMSL metallicities, normalized by the quadrature sum of their respective uncertainties. For the Hypatia uncertainties, we used the standard deviation of all listed values. The gray shaded box highlights the region of $2\sigma$ consistency.}
    \label{fig:metallicity_one_to_one}
\end{figure}

Notably, UV radiation from the host star significantly influences the photochemistry of planetary atmospheres, which in turn affects surface UV environments that are connected to the origin and evolution of life \citep[e.g.,][]{Segura+2003, Buccino+2007}. Depending on the incident UV flux, terrestrial planets orbiting different types of stars are also expected to show different atmospheric features that indicate habitability on Earth, such as $\rm H_2 O$, $\rm C H_4$, $\rm O_3$ and $\rm N_2 O$ \citep{Segura+2003, Kaltenegger+2007, Rugheimer+2015_mstars}. The effects of UV flux on surface and atmospheric conditions also change throughout geological time as chemical evolution modifies concentrations of ozone, methane, and hazes in the atmosphere \citep{Rugheimer+2015}.

On the other hand, the mid-IR portion of host star SEDs may have observational consequences for high-contrast direct imaging of habitable worlds. Significant mid-IR excess, for example, may indicate the presence of exozodiacal dust or warm debris disks \citep[e.g.,][]{Trilling+2008}, which would degrade the quality of direct imaging and spectroscopy of HZ planets. \citet{Mamajek_Stapelfeldt_2023} have indicated in the EMSL which systems have previous detections of circumstellar debris disks, ranking stars with no known dust disks
of any kind in tier A, stars with cold Kuiper Belt (KB) disks that were very optically thin ($L_{\rm IR}/L_\star < 10^{-4}$) in tier B, and stars with any other type of disk in tier C. For completeness, we have added IR fluxes to our catalog where available, which we hope will assist in future analyses of circumstellar debris disks for this sample.

We note that currently only 33 of the stars in our sample have at least one robust space-based UV measurement, and 40 have at least one mid-IR measurement (assuming the quality cuts described in Section \ref{sec:data}). These targets account for just 20-24\% of the catalog. While UV observations of potential HWO targets are being worked on by the NASA Astrophysics Decadal Survey Precursor Science Program (ADSPSP), further work is needed to study the radiation environments of stars in our catalog more closely, particularly for stars currently lacking reliable mid-IR and UV measurements.

\subsection{Stellar composition} \label{subsec:stellar_compositions}

While the EMSL catalog compiled stellar iron abundances from various sources, we have expanded this to include the averaged abundances for 14 elements from the Hypatia catalog \citep{Hinkel+2014_AJ} to facilitate future detailed modeling of planet formation for promising HWO targets. Our catalog includes the primary building blocks of terrestrial planets (e.g., Si, Mg, Al, Fe, and Ni) and the essential ingredients for life as we know it (e.g., C, N, O, P,  and S). We identified 155 ($>94\%$) of our stars with at least one Fe measurement in the Hypatia Catalog, and found measurements of Na, Mg, Al, Si, Ca, and Ni for at least 90\% our sample. The abundances are summarized in Table \ref{tab:abundance_summary} and Figure \ref{fig:hypatia_abundance_histograms}.

\begin{figure}[t!]
    \centering
    \includegraphics[width=0.45\textwidth]{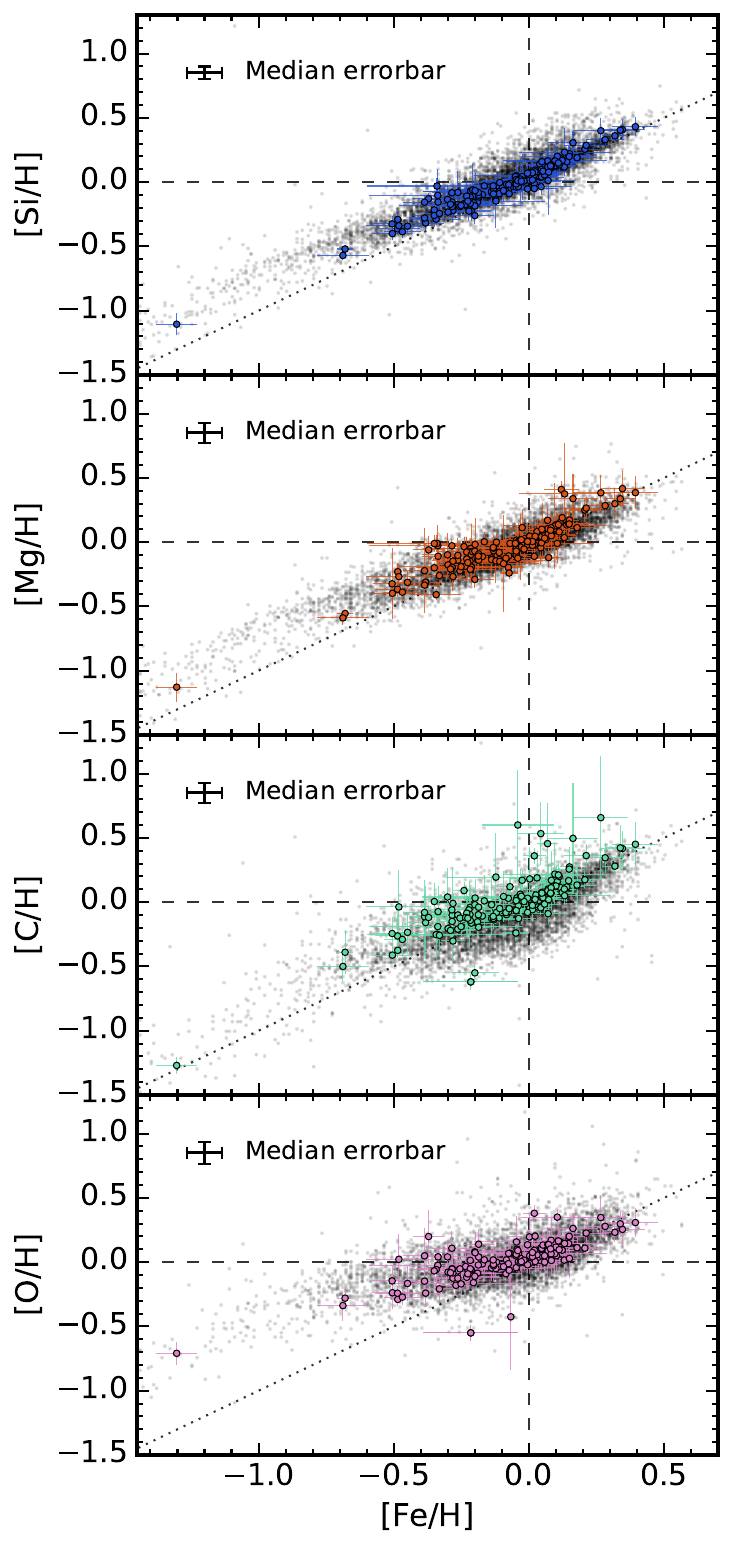}
    \caption{Abundances of silicon, magnesium, carbon, and oxygen vs.~iron from the Hypatia catalog \citep{Hinkel+2014_AJ}. Stars from the catalog presented in this work are shown in color. For reference, all stars with abundance measurements in the Hypatia catalog are plotted as small gray points. The vertical and horizontal dashed gray lines show solar values, and the dotted line shows a 1-to-1 correspondence. A representative median errorbar is shown in the upper left-hand corner of each subplot.}
    \label{fig:abundances}
\end{figure}

Host star composition is directly linked to the protoplanetary disk, which in turn controls the formation and subsequent composition of planetary building blocks \citep[e.g.,][]{Gaspar+2016, Santos+2017, Adibekyan+2021, Cabral+2023}. It is therefore commonly assumed that exoplanet composition mirrors that of the host star \citep[although this is not always the case; see, e.g.,][]{Plotnykov+2020, Schulze+2021}. While the iron abundance [Fe/H] is usually known for nearby stars and often used as a proxy for scaling other stellar abundances, this information alone is insufficient for detailed interior modeling of terrestrial planets because planetary building blocks like Si, Mg, and O are understood not to scale linearly with Fe  \citep[e.g.,][]{Bitsch+2020, Jorge+2022}. Figure \ref{fig:abundances} demonstrates the known non-linearity of the correlation between [Fe/H] and Si, Mg, C, and O abundances in our sample.

Recent studies have implemented devolatilization models, using the detailed compositions of host stars to infer the composition and internal structure of terrestrial planets and thus demonstrating a range of plausible exo-Earth geologies \citep[e.g.,][]{Hinkel+2018, Wang+2022, Unterborn+2023}. For example, solid planet building blocks formed exterior to the water ice line depend strongly on the C/O ratio, with higher C/O in solids leading to smaller water ice fractions as more O is partitioned in gaseous CO and $\rm CO_2$ \citep{Bitsch+2020}.

\begin{figure}[t!]
    \centering
    \includegraphics[width=0.45\textwidth]{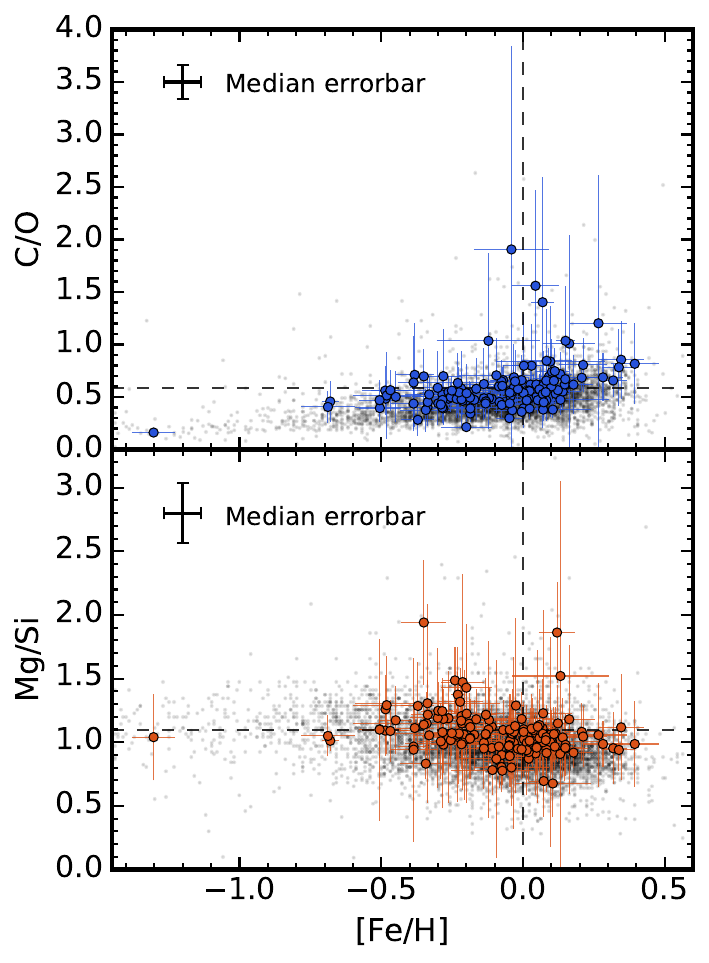}
    \caption{C/O (top) and Si/Mg (bottom) ratios vs.~stellar metallicity [Fe/H] from the Hypatia catalog \citep{Hinkel+2014_AJ}. Stars from the catalog presented in this work are shown in color, and all stars with abundance measurements in the Hypatia catalog are plotted as small gray points for reference. Solar metallicity and abundance ratios \citep[${\rm C/O}=0.59$, ${\rm Si/Mg}=1.10$;][]{Asplund+2021} are indicated by the vertical and horizontal dashed gray lines. A representative median errorbar is shown in the upper left-hand corner of each subplot.}
    \label{fig:CO+SiMg_ratios}
\end{figure}

The Mg/Si ratio is also important to the mineralogy of planet building blocks. According to geochemical models, planets with low Mg/Si ratios tend to cool slowly because their mantle viscosity is high and may therefore quickly lose their volatiles \citep{Spaargaren+2020}. Additionally, different Mg/Si ratios shift the condensation sequence of solid planet building blocks, with lower Mg/Si ratios leading to mechanically stronger mantles and higher Mg/Si leading to weaker mantles \citep{Jorge+2022, Spaargaren+2023}. This can carry important consequences for lithosphere dynamics (i.e., plate tectonics, volcanism, etc.), which can drive different evolutionary pathways that affect habitability. The C/O ratio of our sample ranges from 0.16 to 1.91 \citep[solar ${\rm C/O} = 0.59$;][]{Asplund+2021} and the Mg/Si ratio ranges from 0.68 to 1.94 \citep[solar ${\rm Mg/Si} = 1.10$;][]{Asplund+2021}. Both C/O and Mg/Si as a function of stellar metallicity are shown in Figure \ref{fig:CO+SiMg_ratios}. 

Moreover, the composition of the protoplanetary disk (and thus the host star) plays a role in determining the resulting system architectures of planet formation \citep[e.g.,][]{Adibekyan+2021}. For example, early spectroscopy studies of exoplanet hosts demonstrated that there is a rise in the fraction of stars with planets above solar metallicity \citep{Fischer+2005}. More recently, \citet{Tautvaisiene+2022} analyzed 25 bright planet-hosts from \tess~and showed that stars with high-mass planets tend to be more metal-rich than stars with low-mass planets. The HZ is thought to depend on planetary mass, with larger planets having wider HZs due to their smaller H$_2$O atmospheric column depths \citep{Kopparapu+2014}. Therefore, if metal-rich stars tend to have more massive terrestrial planets, then their HZs may extend further out. This may be an important factor to consider in expanding the list of potential HWO targets.

Another consideration linked to planetary composition is the prevalence of multi-planet systems around stars of different metallicities. Compact multi-planet systems occur more frequently around stars of increasingly lower metallicities, and these systems are nearly mutually exclusive with systems containing hot Jupiters \citep[e.g.,][]{Brewer+2018}. These system architectures may complicate the dynamical stability of otherwise plausible exo-Earths, possibly excluding stable HZ orbits altogether for some systems. The star Groombridge 1830 (CF~UMa), which stands out in our catalog with an extremely low metallicity of ${\rm [Fe/H]} =-1.3$ (labeled in Figure \ref{fig:metallicity_one_to_one}), may therefore be a particularly interesting target. Groombridge 1830 is an early K~dwarf star with no known planetary companions and a predicted EEID angular separation of roughly 50\,mas.

The abundance of bio-essential elements present in the protoplanetary disk may also influence planetary habitability. The element phosphorus, for example, is necessary for all life on Earth because it forms the backbone of DNA and RNA molecules. However, if terrestrial planets form in disks with substantially low P abundances, strong partitioning of P into to core could rule out sufficient bio-available P on the surface \citep{Hinkel+2020}. Only 11 stars in our catalog ($\sim$6.7\%) have measured P abundances, commensurate with the broader dearth of stars with P measurements in the literature---observations are challenging because the typical P absorption lines fall outside of the optical band typically used for spectroscopy \citep{Hinkel+2020}. This motivates future investigations dedicated to measuring P abundances of promising HWO targets.

We note as a caution to the reader that some abundances in the Hypatia Catalog are highly uncertain, with large differences between different catalogs. We found that 12 stars in our catalog have an [Fe/H] spread greater than 0.5\,dex, while 2 stars, HD~38858 and $\delta$ Eri (HD~22049), have an [Fe/H] spread greater than 1.0\,dex. In other cases, the metallicity reported in the EMSL disagrees significantly with the average Hypatia [Fe/H] abundance. For the most significant discrepancy for the star GJ~380 (HD~88230; labeled in Figure \ref{fig:metallicity_one_to_one}), the EMSL [Fe/H] value was $0.21 \pm 0.06$ \citep{Soubiran+2022}, while the mean Hypatia [Fe/H] value was $-0.067$ \citep{Hinkel+2014_AJ}. We ruled out bad crossmatching as an explanation for this discrepancy, and we verified that the range of Hypatia [Fe/H] values spans from $-0.08$ \citep{Woolf+2005} to $-0.05$ \citep{Maas+2016}. This suggests a possible typographical or crossmatching error with the \citet{Soubiran+2022} catalog implemented in the original EMSL for this particular star.

We urge future studies using our catalog to exercise caution by checking the span of abundance measurements before assuming a particular value. In our catalog, we include the minimum and maximum Hypatia abundance values along with their literature sources to facilitate this check.

\subsection{Stellar environments} \label{subsec:stellar_environments}

We have identified 108 stars in our sample that exhibit variability, flares, and/or X-ray emission. Furthermore, the EMSL contains over 100 stars in binary systems and 4 post-main-sequence stars ($\log (g) < 4$). The stars are mostly intermediate in mass, with 66 F-type, 55 G-type, 40 K-type, and only 3 M-type stars. While this is by no means an exhaustive list of factors that may affect habitability, it provides a useful starting point for discussing how past, present, and future stellar environments are intertwined with planetary surface conditions necessary for life.

Periodic stellar variability is often an indication of radial pulsations (typically in evolved massive stars) or rotationally-modulated brightness differences caused by photospheric heterogeneities (typically in cooler dwarf stars). In the latter case, analysis of the photometric time series can yield constraints on the rotation period and star spot coverage of the stellar photosphere. Stellar rotation has been theoretically shown to affect the limits of the HZ, especially in rapidly rotating A and F stars where the rotational forces cause the star to become oblate and develop a pole-to-equator temperature gradient \citep{Ahlers+2022}. Star spot coverage on the other hand is linked to magnetic fields, and hence stellar activity, which has been extensively studied in the context of M-dwarf planet habitability \citep[e.g.,][]{Joshi+1997, Scalo+2007, Gallet+2017, Gunther+2020}. Stellar flares, which are associated with high stellar activity, are high-energy transients that can emit large quantities of UV and X-ray radiation, especially in M and K dwarf stars. Flares present challenges to potential habitability because they can alter the atmospheric chemistry of terrestrial planets (primarily generating $\rm O_3$ and $\rm N_2 O$) and expose their surfaces to high levels of ionizing radiation \citep[e.g.,][]{Ridgway+2023}.

Stellar age can also present challenges to potential habitability. While the majority of stars in our sample are main-sequence stars, there are a handful of subgiant stars---notably, $\delta$ Eri (GJ~150) has the lowest surface gravity with $\log (g) = 3.78$. The HZs of post-main sequence stars move outward as the stellar luminosity increases, which can make them more accessible to direct-imaging surveys. However, stellar mass loss during the RGB and AGB phases of post-main sequence evolution can destabilize planetary orbits and drive strong stellar winds that work to erode planetary atmospheres, possibly reducing the chances of habitability \citep[e.g.,][]{Ramirez+2016}. More detailed modeling is needed to assess the importance of these effects for the few subgiant stars in our catalog.

Ultimately, a holistic assessment of each system will be necessary to select the most promising targets for HWO, informed by both observability (limited by technology) and likely scientific return (limited by precursor knowledge). Therefore, we do not attempt to narrow down the list of potential targets in this work, but rather present what is known about each system at face value in hopes that this information will be used holistically in future studies.

\subsection{Data availability} \label{subsec:data_availability}

Our catalog of 164 stars is publicly available to the community via the authors' website\footnote{\url{https://sites.google.com/berkeley.edu/spores-hwo}} and the online journal. A list of columns included in the current version of the catalog and their descriptions are provided in Table \ref{tab:columns} in Appendix \ref{app:extras}. We invite the HWO community to use this catalog as a resource for their own studies.

Finally, we intend to update this catalog with additional stars and properties at a future time. Expanding the sample of potential HWO targets will enable a more comprehensive investigation in mission design trade space, as a significant portion of EMSL targets may ultimately be determined to be unsuitable for the HWO planet survey \citep[see, e.g.,][]{Tuchow+2024}.

\section{Conclusion} \label{sec:conclusion}

We have created a catalog of 164 stars that are promising targets for the Habitable Worlds Observatory, a future $\sim$6-meter class space mission capable of high-contrast direct imaging and spectroscopy \citep{astro2020}. Our catalog derives from the NASA Exoplanet Exploration Program's Mission Star List for HWO, which selected 164 nearby stars whose hypothetical exo-Earths would be most accessible to a high-contrast direct-imaging survey in terms of angular separation and planet-star contrast ratio \citep{Mamajek_Stapelfeldt_2023}. We include over 1700 individual stellar abundance measurements for 14 elements, over 900 photometry measurements spanning from 151.6 $\rm nm$ to 22 $\rm \mu m$, flare rate estimates for 46 stars, variability metrics for 78 stars, and X-ray detections for 46 stars.

Bayesian analysis of the stellar spectral energy distributions yielded broadly consistent stellar parameters compared to previously published values reported in the EMSL. Iron abundances averaged from the Hypatia Catalog \citep{Hinkel+2014_AJ} are also mostly consistent with those drawn from the EMSL, but often with much more conservative uncertainties due to significant differences between the various Hypatia source catalogs. The SEDs and different compositions of the potential host stars likely have important consequences affecting their HZs and potential habitable planets, which we have discussed in detail.

We emphasize that the catalog and discussion presented here is intended to aid future work investigating these systems more closely. This work is the first in a series of forthcoming precursor science papers that will result from the NASA Astrophysics Decadal Survey Precursor Science (ADSPS) program\footnote{Funded through ADSPS22 Grant Number 80NSSC23K1476.}. Subsequent precursor science work is critical to HWO trade studies that will determine the final mission design. Throughout this work, we have hinted at future avenues toward science with HWO, including closer investigations of the stars' high-energy and mid-IR radiation environments, more complete surveys of bio-essential element abundances, and studies of dynamical stability in systems with previously detected planets or a higher likelihood of being a compact multi-planet system. 

Additionally, future work should include determining the depth of past searches for planets in these systems using archival photometry and public RV observations \citep[e.g.,][]{Howard+2016, Laliotis+2023}. A follow-up investigation of stellar variability and rotation rates would also be useful for accurately measuring stellar ages with gyrochronology, and as a proxy for stellar activity. Lastly, as future studies begin to winnow down the list of possible HWO targets, additional stars should also be added to the list to allow for more flexibility in mission design trades. Investing in this work early on in the mission development will reduce costs while maximizing the likelihood of reaching HWO's goal of detecting life or placing meaningful constraints on the rarity of life on other planets. The EMSL sample studied here was compiled using a somewhat limiting definition of the HZ and the assumption of only circular orbits, but this can be expanded by extending the permissible conditions for habitability and adding planets with eccentric orbits. These future pathways toward planet properties were discussed in our ADSPS program proposal.

Our catalog is publicly available, and we invite other members of the HWO community to build off of our work in future studies leading up to HWO and beyond. Ultimately, the subsequent work that follows from this catalog will play a critical role in the success of HWO and constraining the existence of life beyond the Solar System.

\section*{Acknowledgments}
We acknowledge support from the NASA Astrophysics Decadal Survey Precursor Science (ADSPS) program under Grant Number 80NSSC23K1476.
C.K.H.~acknowledges support from the National Science Foundation (NSF) Graduate Research Fellowship Program (GRFP) under Grant No.~DGE 2146752.
B.A.A.~acknowledges support from the Cal-Bridge Summer Research Program.
This research has made use of the NASA Exoplanet Archive, which is operated by the California Institute of Technology, under contract with the National Aeronautics and Space Administration under the Exoplanet Exploration Program.
The research shown here acknowledges use of the Hypatia Catalog Database, an online compilation of stellar abundance data as described in \citet{Hinkel+2014_AJ}, which was supported by NASA's Nexus for Exoplanet System Science (NExSS) research coordination network and the Vanderbilt Initiative in Data-Intensive Astrophysics (VIDA).
This work has made use of data from the European Space Agency (ESA) mission
{\it Gaia} (\url{https://www.cosmos.esa.int/gaia}), processed by the {\it Gaia}
Data Processing and Analysis Consortium (DPAC,
\url{https://www.cosmos.esa.int/web/gaia/dpac/consortium}). Funding for the DPAC
has been provided by national institutions, in particular the institutions
participating in the {\it Gaia} Multilateral Agreement.
This research has made use of the Washington Double Star Catalog maintained at the U.S. Naval Observatory.
This research made use of Lightkurve, a Python package for Kepler and TESS data analysis \citep{lightkurve}.
This research has made use of the International Variable Star Index (VSX) database, operated at AAVSO, Cambridge, Massachusetts, USA.

\facilities{
    ADS,
    CDS,
    Exoplanet Archive,
    Gaia,
    GALEX,
    HIPPARCOS,
    MAST,
    TESS,
    WISE
}

\software{
    ARIADNE \citep{Vines+2022MNRAS},
    astropy \citep{astropy_2013_A&A, astropy_2018_AJ},
    bokeh \citep{bokeh},
    corner \citep{corner},
    dynesty \citep{Speagle2019},
    IPython \citep{Perez+2007},
    isochrones \citep{isochrones},
    Lightkurve \citep{lightkurve},
    matplotlib \citep{Hunter+2007},
    NumPy \citep{vanderWalt+2011, Harris+2020},
    Pandas \citep{mckinney-proc-scipy-2010, reback2020pandas}
}

\appendix

\section{Modifications to the EMSL}\label{app:alterations}
Our catalog contains most of the original information included in the EMSL. However, we identified a few errors or inconsistencies that we have corrected here. These are summarized as follows.
\begin{itemize}
    \item The EMSL contained an empty column (\texttt{eyes\_lum}), which we have removed from our catalog.
    \item The star Toliman ($\alpha$~Cen~B; HIP~71681) was originally flagged as a planet host in the EMSL, which we have changed here. While an Earth-mass planet candidate orbiting $\alpha$~Cen~B was initially detected in radial velocity measurements of the system \citep{Dumusque+2012}, subsequent analysis of the same data attributed the RV signal to the observation window function, thus ruling out $\alpha$~Cen~Bb as a bona fide planet \citep{Rajpaul+2016}. For this star, we have changed the \texttt{sy\_pnum} column to 0 and the \texttt{sy\_planets\_flag} column to ``N.''
    \item \citet{Mamajek_Stapelfeldt_2023} noted several stars in the EMSL that were previously reported to be binary, but are likely spurious (see their Appendix E). For example, the star 104~Tau (HIP~23835; WDS~J05074$+$1839) was labeled as a ``Dubious Double'' in the Washington Double Star (WDS) Catalog notes, which could be attributed to a plate flaw, a positional typo in the original publication, or an optical double disappearing due to drastically different proper motions between the components \citep{Mason+2001}. Lacaille~9352 (GJ~887; HIP~114046; WDS~J23059$-$3551) was also indicated as a binary in the WDS Catalog, but subsequent RV measurements ruled out a companion star while confirming two super-Earths in the system \citep{Jeffers+2020}. HD~4628 (HIP~3765; WDS~J00484$+$0517), HD~5015 (HIP~4151; WDS~J00531$+$6107), $\zeta$~2~Ret (HIP~15371; WDS~J03182$-$6230), Chara ($\beta$~CVn; HIP~61317; WDS~J12337$+$4121), and 58~Oph (HIP~86736; WDS~J17434$-$2141) also had some previous indication of binarity in the literature, but were subsequently ruled out either by \citet{Mamajek_Stapelfeldt_2023} or previous authors. For each of these stars, we have removed the spurious data from the \texttt{wds\_sep} and \texttt{wds\_deltamag} columns.
    \item While the original EMSL included apparent $V$ magnitudes and $B-V$ color indices, for convenience we have added a column with the isolated $B$ magnitude. We also add a column for the $B$ magnitude uncertainty, which we calculated by adding the $V$ magnitude uncertainty and $B-V$ color uncertainty in quadrature. 
\end{itemize}

\section{Additional figures and tables}\label{app:extras}

This section contains additional figures and tables referenced in the main text.

\begin{figure*}[t!]
    \centering
    \includegraphics[width=0.8\textwidth]{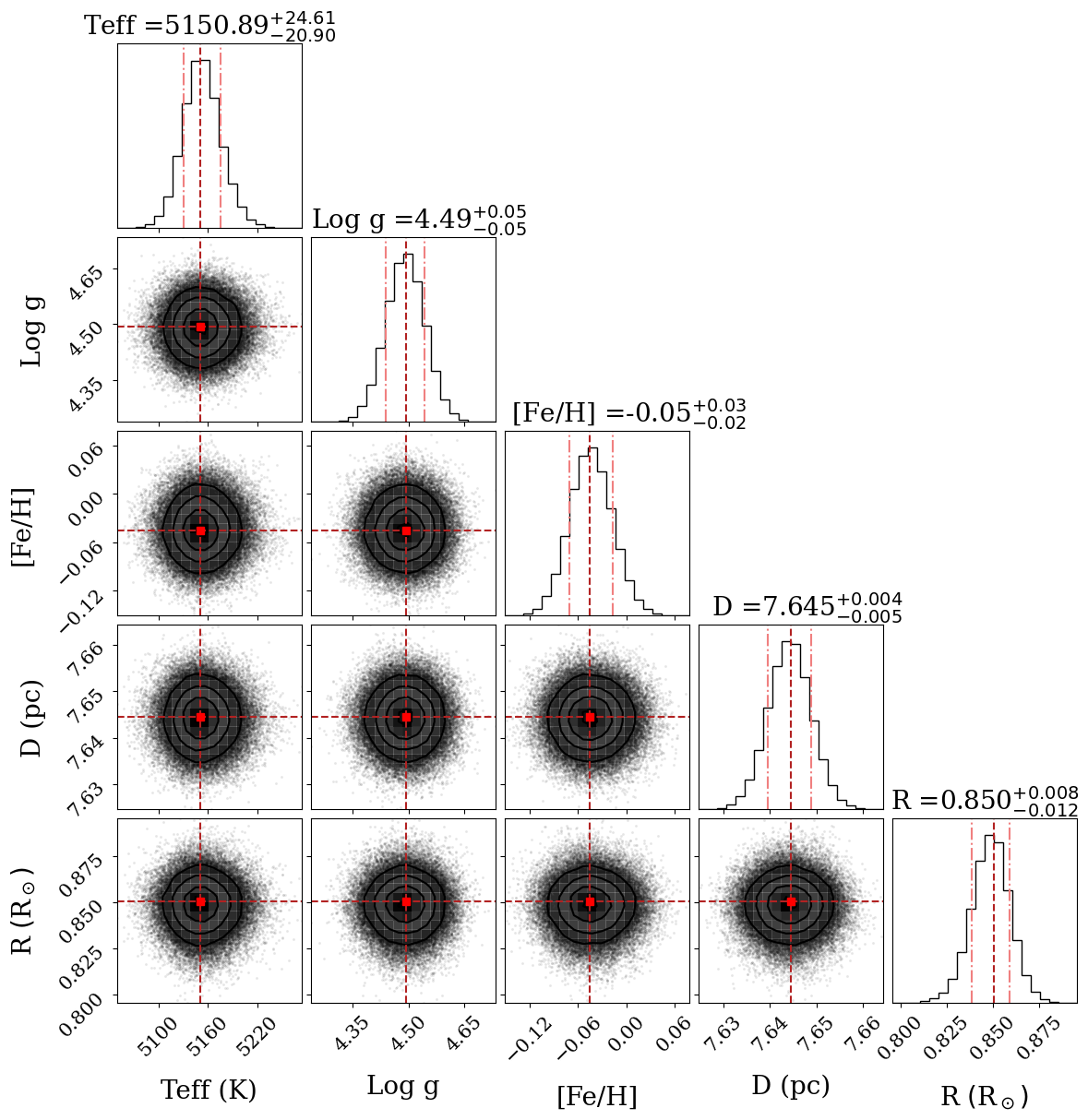}
    \caption{Posterior distributions derived from the best-fit SED model for HD~10476 shown in Figure \ref{fig:example_sed}. The red dots show the median posterior values, while the vertical red lines in the marginalized posterior distributions show the 16th, 50th, and 84th percentile values.}
    \label{fig:sed_posteriors}
\end{figure*}

\clearpage\newpage
\bibliography{new.ms}{}
\bibliographystyle{aasjournal}

\startlongtable
\begin{deluxetable*}{chcV{8cm}cV{2cm}}
\tablecaption{Column definitions. \label{tab:columns}}
\tablehead{
\colhead{Column name} & \nocolhead{OldName} & \colhead{Unit} & \colhead{Description} & \colhead{EMSL} & \colhead{Ref.~\#}}
\startdata
\texttt{   TIC            }                &    \texttt{tic\_id}                                     &                     &   TIC designation in the TESS Input Catalog.                                                                                                                                                                                                                                                                                      [tic\_id]   &   Y  & 86 \\ \hline
\texttt{   HIP            }                &    \texttt{hip\_name}                                   &                     &   HIP designation in the Hipparcos Catalog.                                                                                                                                                                                                                                                                                       [hip\_name]   &   Y  & 26 \\ \hline
\texttt{   HIPComp        }                &    \texttt{hip\_compname}                               &                     &   Component letters provided for when multiple stars are associated with the HIP entry (following CCDM, WDS, or SIMBAD).                                                                                                                                                                                                          [hip\_compname]   &   Y  & 62 \\ \hline
\texttt{   HD             }                &    \texttt{hd\_name}                                    &                     &   HD designation in the Henry Draper Catalog.                                                                                                                                                                                                                                                                                     [hd\_name]   &   Y  & 12 \\
\enddata
\end{deluxetable*}

\end{document}